\documentclass[aps,pra,reprint,showpacs,floatfix,superscriptaddress,final]{revtex4-1}
\usepackage{amsmath,amssymb}
\usepackage[utf8]{inputenc}
\usepackage{graphicx}
\usepackage{color,soul}
\usepackage[colorlinks=true,linkcolor=blue,urlcolor=blue,citecolor=blue]{hyperref}
\usepackage{bm}

\def \hat {{}}

\newcommand*{\av}[1]{\left\langle#1\right\rangle}
\newcommand*{\fourier}[1]{\mathcal{F}\mathopen{}\mathclose{}\bgroup\left[#1\aftergroup\egroup\right]}

\newcommand*{\ket}[1]{\left|#1\right\rangle}
\newcommand*{\ve}{\varepsilon}
\newcommand*{\sfrac}[2]{{\textstyle\frac{#1}{#2}\displaystyle}}
\newcommand*{\yd}{\dagger}
\newcommand*{\nd}{{\vphantom{\dagger}}}
\renewcommand*{\Im}{\operatorname{Im}}
\renewcommand*{\Re}{\operatorname{Re}}
\newcommand*{\gammau}{\gamma_\uparrow}
\newcommand*{\gammad}{\gamma_\downarrow }
\newcommand*{\gammap}{\gamma_\downarrow + \gamma_\uparrow}

 \global\long\def\arr4#1#2#3#4{\left(\begin{array}{cc} #1 & #2 \\ #3 & #4\\ \end{array}\right)}

\def\clap#1{\hbox to 0pt{\hss#1\hss}}

\def\mathclap{\mathpalette\mathclapinternal}

\def\mathclapinternal#1#2{%
	\clap{$\mathsurround=0pt#1{#2}$}}

\begin{document}

\title{Fermionic formalism for driven-dissipative multi-level systems}

\author{Yulia Shchadilova}
\thanks{These authors contributed equally to this work.}
\affiliation{Department of Physics, Harvard University, Cambridge, Massachusetts 02138, USA}

\author{Mor M. Roses}
\thanks{These authors contributed equally to this work.}
\affiliation{Department of Physics, Bar Ilan University, Ramat Gan 5290002, Israel}

\author{Emanuele G. Dalla Torre}
\affiliation{Department of Physics, Bar Ilan University, Ramat Gan 5290002, Israel}

\author{Mikhail D. Lukin}
\affiliation{Department of Physics, Harvard University, Cambridge, Massachusetts 02138, USA}

\author{Eugene Demler}
\email{demler@physics.harvard.edu}
\affiliation{Department of Physics, Harvard University, Cambridge, Massachusetts 02138, USA}

\begin{abstract}
We present a {\it fermionic} description of non-equilibrium multi-level systems.
Our approach uses the Keldysh path integral formalism and allows us to take into account periodic drives, as well as dissipative channels.
The technique is based on the Majorana fermion representation of spin-1/2 models which follows earlier applications in the context of spin and Kondo systems.
We apply this formalism to problems of increasing complexity: a dissipative two-level system, a driven-dissipative multi-level atom, and a generalized Dicke model describing many multi-level atoms coupled to a single cavity. We compare our theoretical predictions with recent QED experiments and point out the features of a counter-lasing transition. Our technique provides a convenient and powerful framework for analyzing driven-dissipative quantum systems, complementary to other approaches based on the solution of Lindblad master equations.
\end{abstract}


\maketitle

\section{Introduction}

Driven dissipative many-body systems are the subject of current experimental and theoretical investigations at the interface of condensed matter physics and quantum optics. In these systems, the interplay between unitary dynamics and dissipative channels may lead to non-equilibrium steady states with properties substantially different from quantum phases in thermal equilibrium. A recent example of such a system involves pump-probe experiments in which driven, out-of-equilibrium phonons give rise {to} superconducting correlations at room-temperature~\cite{fausti2011light, Mitrano2016, knap2016dynamical, sentef2016theory, babadi2017theory}. Atomic and molecular systems offer another well-known example of driven-dissipative systems. Here, the interplay between driving, dissipation, and interaction facilitates observations of phase transitions such as the superradiant phase transition that are hard to explore in equilibrium setups (see Ref.~\cite{kirton2018introduction}
for an introduction).

Understanding non-equilibrium phase transitions of open quantum systems is a challenging theoretical problem.  While a number of powerful theoretical tools has been developed for the description of equilibrium phase transitions~\cite{sachdev2007quantum}, fewer tools are available for non-equilibrium problems. {The analysis of driven-dissipative systems requires mathematical tools and approximation schemes which treat the collective behavior of large ensembles, strong correlations, and non-equilibrium physics on an equal footing. In the field of quantum optics, master equation approaches are commonly used since they are well suited to work with these types of systems~\cite{scully1999quantum}. However, alternative approaches can provide new insights, using analogies with out-of-equilibrium solid state systems.}

Field-theoretical approaches, widely used in condensed matter and high-energy physics, were developed to describe out-of-equilibrium many-body systems~\cite{Keldysh1965,Keldysh2003,tsvelik2007quantum,kamenev2011field}. In fact, recent theoretical papers demonstrated the particular strength of the Keldysh approach for the description of non-equilibrium phase transitions in open quantum optical systems ~\cite{szymanska2006,DallaTorre2013,buchhold2013dicke,sieberer2013dynamical,lang2016critical,DallaTorre2016,marino2016quantum,sieberer2016keldysh}. These works adopted a {\it bosonic} approach, where the continuum limit of a spin model was considered. Here, we instead opt for a {\it fermionic} approach, which enables us to describe systems with a finite number of allowed states, such the lambda or $W$ schemes. This approach allows us to study the non-equilibrium steady states induced by the interplay between periodically driven fields and dissipative channels. 

\begin{figure*}[t]
	\begin{centering}
		\includegraphics[width=0.8\linewidth]{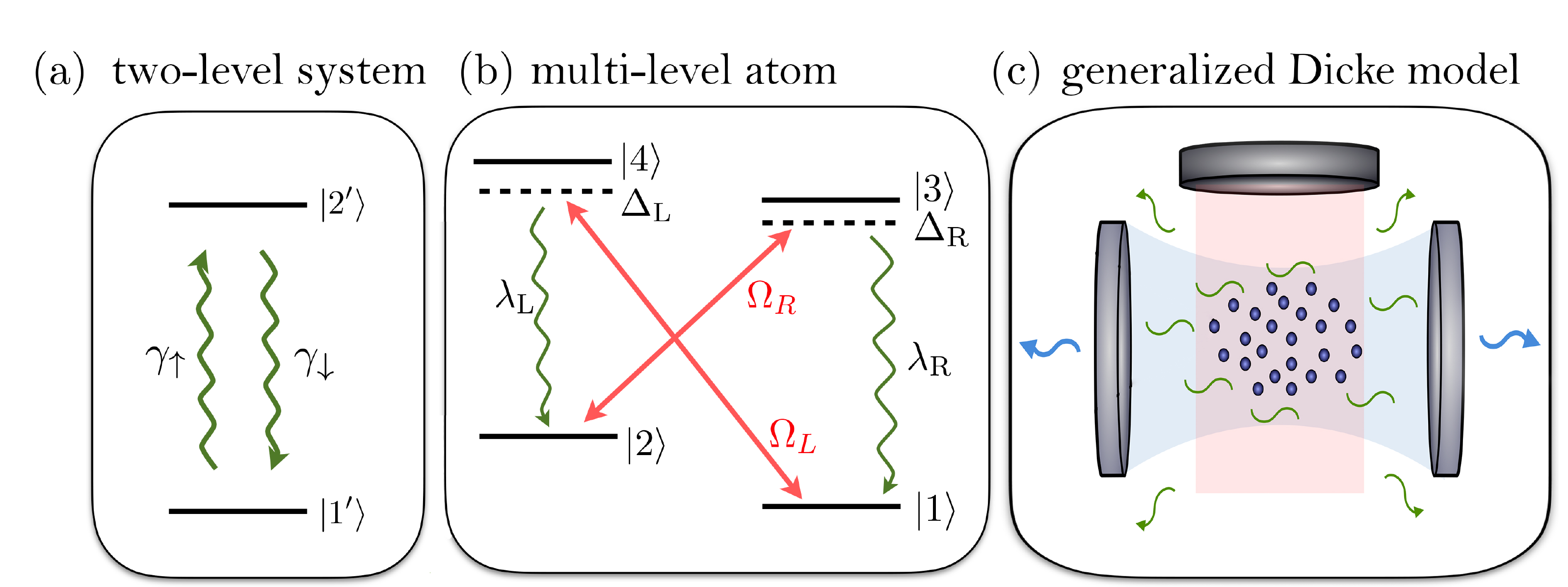}
		\par\end{centering}
	\protect\protect
	\caption{%
		\label{fig:Sketch_Spin}%
			Sketch of the open quantum systems described in this paper: (a) The two-level system examined in Sec.~\ref{sec:sp}, coupled by dissipative channels (green wiggly arrows). (b) The 4-level atom (W scheme) considered in Sec.~\ref{sec:model}, including pumping fields (red arrows) and dissipative processes.
			(c) The Dicke model considered in Sec.~\ref{sec:phasetr}, describing the coupling between many driven-dissipative atoms and the quantized field of an optical cavity.
	}

\end{figure*}

In this paper we show how to use fermionic path integrals to describe open quantum systems of increasing complexity (see Fig.~\ref{fig:Sketch_Spin}). First, in Sec.~\ref{sec:sp} we  consider a two-level system (spin-$\frac12$) with dissipation.
We use the `drone'-fermion approach to convert a single spin to two fermions, a Dirac (complex) and a Majorana fermion.
\cite{martin1959,mattis1965theory,Kenan1966,Spencer1967,spencer1968theory,casalbuoni,berezin1977particle,Tsvelik1992,Coleman1993,Shastry1997,Shnirman2003}. This approach allows us to construct diagrammatic techniques for the description of the steady-state properties of the system.
The Majorana representation has several advantages and does not require any constraints, in contrast to bilinear forms of fermions~\cite{abrikosov1965anomalous} and bosons (i.e., Schwinger-boson representation)~\cite{tsvelik2007quantum}, which require the imposition of constraints onto the Hilbert space.
Another advantage of the Majorana representation is that calculations of spin-spin correlation and response functions can be simplified~\cite{Shnirman2003,mao2003spin,Biswas2011,Schad2015}. Next, in Sec.~\ref{sec:dis} we move the case of a multi-level system, namely a driven-dissipative 4 level scheme. Using fermionic path integrals, we determine the conditions under which this system can be effectively described as a two-level system. In Sec.~\ref{sec:phasetr}, we move to a yet higher degree of complexity. We consider the coupling between many atoms and a single cavity mode, giving rise to a driven-dissipative generalized Dicke model.
Our findings are relevant to a large number of different experimental implementations, including cavity QED ~\cite{black2003observation,baumann2010dicke,Baumann2011,brennecke2013real,klinder2015dynamical,klinder2015observation,
Zhiqiang2017,zhiqiang2018dicke},  trapped ion~\cite{genway2014}, and superconducting circuits (see Ref.~\cite{kirton2018introduction} and references therein). For concreteness, we focus on a  recent realization of the Dicke model~\cite{Dimer2007,Zhiqiang2017}, where we find signatures of an unusual lasing instability. Finally, in Sec.~\ref{sec:Lindblad} we compare our results with a mean-field approximation to the Lindblad master equation, giving rise to a Maxwell-Bloch description of the system~\cite{Kirton2017}.

\section{Dissipative two level system}
\label{sec:sp}

\subsection{Majorana-Dirac fermion representation}
\label{sec:maj}

In this section we provide a \emph{fermionic} description of a single spin-$\frac12$ coupled to two Markovian baths.
This system can be described by the following Hamiltonian:
\begin{align}
	\label{eq:sp12-ham}
	\begin{split}
		H=&\omega^\nd_zS^z+\sum\limits_{\mathclap{k,\sigma={L,R}}}\nu^\nd_{k\sigma}d^\yd_{k,\sigma}d^\nd_{k,\sigma}
		\\
		&+\sum\limits_k\frac{\Omega_L}{\Delta_L}\lambda_{k,L}\left(d^\nd_{k,L}S^-+d^\yd_{k,L}S^+\right)
		\\&
+\sum\limits_k\frac{\Omega_R}{\Delta_R}\lambda_{k,R}\left(d^\nd_{k,R}S^++d^\yd_{k,R}S^-\right)
	\end{split}
\end{align}
where $\omega_z$ is the two-level splitting, $\nu_{k,\sigma}$ is the bath frequency for a given $k$ and polarization $\sigma$, $\frac{\Omega_\sigma}{\Delta_\sigma}\lambda_{k,\sigma}$ is the two-level to bath coupling coefficient, $d^{(\yd)}_{k,\sigma}$ is the annihilation (creation) operator for a bath photon in a given $k$ and $\sigma$ and $S^\alpha$ are the spin operators satisfying $\left[S^\alpha,S^\beta\right]=i\varepsilon_{\alpha\beta\gamma}S^\gamma$.
The first step would be to describe the two-level system as two fermions $c_1$ and $c_2$, each satisfying $\left[c^\nd_i,c^\yd_j\right]=\delta_{i,j}$.
This notation has the following constraint:
\begin{align}
	\label{eq:const}
	c^\yd_1c^\nd_1+c^\yd_2c^\nd_2=&1,
\end{align}
In this notation the various spin operators ($S^\alpha$) are transformed as:
\begin{align}
	S^z=&c^\yd_2c^\nd_2,&
	S^+=&c^\yd_2c^\nd_1=\left(S^-\right)^\yd.
\end{align}
Solving this system \eqref{eq:sp12-ham} with the constraint \eqref{eq:const} using field-theoretical tools can become cumbersome.

We can simplify the problem by using the Majorana representation in which the constraint is automatically fulfilled. To achieve this goal, we represent the complex fermions $\hat c_1$ and $\hat c_2$ as a linear combination of
four Majorana fermions
\begin{eqnarray}\label{eq:majo}
\hat c_1  &=&\frac{1}{2}\left(\hat  \eta_z+i\hat \eta_0\right), \\ \nonumber
\hat  c_2 &=&\frac{1}{2}\left(\hat  \eta_x+i\hat \eta_y\right).
\end{eqnarray}
Here the Majorana fermions $\hat \eta_i$ satisfy $ \hat \eta_i=\hat \eta_i^\dagger$, $\hat \eta_i^2=1$, and $\lbrace \hat \eta_i,\hat \eta_j \rbrace =2\delta_{ij}$  for all  $i\in \left[0,x,y,z\right]$.

By using the Majorana representation~\eqref{eq:majo} , we can express the constraint~\eqref{eq:const} as
\begin{equation}\label{eq:majo-const}
\hat \eta_z\hat \eta_0+\hat \eta_x\hat \eta_y=0.
\end{equation}
Equation \eqref{eq:majo-const} implies that in the physical space {$ \hat \eta_z \hat \eta_0\ket\psi = -\hat \eta_x \hat \eta_y\ket\psi$}.
By multiplying both sides of the equation by $\hat \eta_z$ (from the left) and using the properties
of Majorana fermions, we obtain:
\begin{equation} \label{eq:left}
          \hat \eta_0\ket\psi= -  \hat \eta_x  \hat \eta_y \hat \eta_z  \ket\psi.
\end{equation}
Equation \eqref{eq:left} can now be used to eliminate the Majorana fermion {$\hat \eta_0$} from the
Hamiltonian {\eqref{eq:sp12-ham}}.

In practice, it is convenient to use the mixed Majorana-Dirac fermion representation (`drone'-fermion)~\cite{Kenan1966,Spencer1967,spencer1968theory, Shnirman2003},  where two Majoranas are combined into a single complex (Dirac) fermion $\hat f=\frac{1}{2}\left(\eta_x + i\eta_y\right)$, and a third Majorana fermion is denoted by $\hat \eta \equiv \hat \eta_z$. Using Eqs.~\eqref{eq:majo} and~\eqref{eq:left}, we express the Hamiltonian~{\eqref{eq:sp12-ham}} through the $\hat \eta$ and $\hat f$ operators
\begin{eqnarray}\label{eq:mixedrepr}
\hat c_2^\dag \hat c_1 &=& \hat f^\dag \hat \eta , \;\;\;\; \hat c_1^\dag \hat c_2 =\hat \eta \hat f \\ \nonumber
\hat c_2^\dag \hat c_2 &=&  \hat f^\dag \hat f, \;\;\;\; \hat  c_1^\dag \hat c_1 = \hat f \hat f^\dag.
\end{eqnarray}

{ The mixed Majorana-Dirac representation can be conveniently mapped into a spin-$\frac{1}{2}$ system~\cite{Kenan1966,Spencer1967,spencer1968theory, Shnirman2003}. To achieve this task, one needs to identify the full and empty states of the Dirac fermion, respectively, with the spin-up and spin-down states of a spin-$\frac{1}{2}$ system. The Majorana fermion is used to fulfill the canonical spin commutation relations. Formally, the mapping is given by}
\begin{eqnarray}\label{eq:spinrepr}
\hat S^x &=& \frac{1}{2} \left( \hat f^\dag - \hat f \right)\hat\eta,   \;\;\;
\hat S^y  = -\frac{i}{2} \left(  \hat f^\dag + \hat f \right) \hat \eta  , \\ \nonumber
\hat S^z &=& f^\dagger f  - \frac12  \;\;\;.
\end{eqnarray}
or equivalently
\begin{eqnarray}\label{eq:spinrepr2}
\hat S^x &=& \frac{1}{2} \left( \hat f^\dag + \hat f \right)\hat\tau_x = \frac{1}{2} \eta_x \tau_x, \\ \nonumber
\hat S^y  &=& -\frac{i}{2} \left(\hat f^\dag - \hat f  \right)  \tau_x = -\frac{1}{2} \eta_y \tau_x , \\ \nonumber
\hat S^z &=& \frac{1}{2} \hat  \eta \hat  \tau_x = \frac{1}{2} \eta_z \tau_x   \;\;\;
\end{eqnarray}
Here $\hat S^\alpha$ are spin-$\frac{1}{2}$ operators satisfying the canonical commutation relations $[\hat S^\alpha , S^\beta] = i \epsilon_{\alpha\beta\gamma} \hat S^\gamma$ where $\epsilon_{\alpha\beta\gamma}$ is the Levi-Civita symbol; and $\hat \tau_x = - i \hat \eta_x \hat \eta_y \hat \eta_z = (1-2 \hat f^\dag \hat f)\hat \eta$.
Note that the ``copy-switching'' operator $\tau_x$ (for discussion see Ref.~\cite{Shnirman2003}) commutes with the Hamiltonian {\eqref{eq:sp12-ham}} and is thus time independent. This property will allows us to simplify the calculation of some spin-spin correlation and response functions ~\cite{Shnirman2003,mao2003spin,Biswas2011,Schad2015}.

\subsection{Dissipation of a single spin}
\label{sec:diagr}

{ Using the mixed Majorana-Dirac representation introduced in the previous section, we can rewrite the effective Hamiltonian~{\eqref{eq:sp12-ham}} as}
\begin{eqnarray} \label{eq:Hamilt_diss}
\hat{\mathcal{H}}_{\rm (0)} & = &     \omega_z \hat f^\dag \hat f +
	\sum_{\mathclap{k,\sigma=\lbrace L,R \rbrace} } \nu^\nd_{k\sigma} \hat d_{k,\sigma}^{\dag}\hat d^\nd_{k,\sigma} \\ \nonumber
&+&    \sum_{k}  \frac{\Omega_{R}}{\Delta_R} \lambda_{k,R} \left( \hat d_{k,R} \hat f^\dag \hat \eta + \hat d_{k,R}^\dag  \hat \eta \hat f\right) \nonumber\\
&+& \sum_{k}  \frac{\Omega_{L}}{\Delta_L}  \lambda_{k,L}  \left( \hat d_{k,L} \hat \eta \hat f  + \hat d_{k,L}^\dag  \hat f^\dag \hat \eta\right) \nonumber
\end{eqnarray}
where the Majorana $\hat \eta$ and Dirac (complex) $\hat f$ fermions are introduced using Eq.~\eqref{eq:mixedrepr}.

We { now} study the properties of the system using the Green's functions on the Keldysh contour~\cite{Keldysh1965,Keldysh2003}. In particular, we are interested in the description of the non-equilibrium steady state which is the result of the interplay between driving and dissipation processes.

\paragraph{Bosonic bath. -- }
We describe the bosonic bath using the greater and lesser Green's functions on the Keldysh contour
\begin{eqnarray}
D_{k,L}^{>}(t,t') =  - i \av{ \hat d_{k,\sigma} (t) \hat d^\dag_{k,\sigma}(t)} \\ \nonumber
D_{k,L}^{<}(t,t') =  - i \av{ \hat d_{k,\sigma}^\dag (t') \hat d_{k,\sigma}(t)}
\end{eqnarray}

In this work, we focus on Markovian baths, characterized by $A_{k,\sigma}(\omega) = S_{k,\sigma}(\omega)$ (see also Sec.~\ref{sec:RWA} below). Because, by definition $D^<_{k,\sigma}(\omega) = A_{k,\sigma}(\omega) - S_{k,\sigma}(\omega)$, lesser Green's functions of Markovian baths are identically equal to zero. Hence, the integral effect of all bosonic modes can be described by introducing the effective parameters $\gamma_\uparrow$ and $\gamma_\downarrow$
\begin{eqnarray} \label{eq:bath_properties1}
&& \sum_{k}\frac{ \Omega_L^2 \lambda_{k,L}^{2} }{8\Delta_L^2 }  D_{k,L}^{>}(\omega)	 =  - i  \gamma_\uparrow, \\ \nonumber
&& \sum_{k}\frac{ \Omega_R^2 \lambda_{k,R}^{2} }{8\Delta_R^2 } D_{k,R}^{>}(\omega) =  - i \gamma_\downarrow, \label{eq:bath_properties3}
\end{eqnarray}

\paragraph{Fermions. -- }
The Green's functions of the $f$ and $\eta$ fermions are defined as
\begin{eqnarray} \label{eq:def_GF}
G^>_\eta(t,t') &=& - i \av{\hat \eta (t) \hat \eta(t')}, \; G^<_\eta(t,t') = i \av{ \hat \eta(t') \hat \eta (t)}\\ \nonumber
G^>_f(t,t') &=& - i \av{\hat f (t) \hat f^\dag(t')}, \; G^<_f(t,t') = i \av{ \hat f^\dag(t') \hat f (t)}
\end{eqnarray}

In a steady state, the Green's functions only depend on the time differences and one can introduce the function $h_f(\omega)$ and the spectral function $\rho_f(\omega)$, such that { the Fourier transformed} Green's functions read~\cite{keldysh1986high,Tikhodeev2008}
\begin{eqnarray}\label{eq:GF_def}
G^>_f(\omega) &=& - i\pi \left( 1 + h_f(\omega)\right)  \rho_f(\omega), \\  \nonumber
G^<_f(\omega) &=&  i\pi \left( 1 - h_f(\omega)\right) \rho_f(\omega)
\end{eqnarray}
where $h_f(\omega)$ is connected to the occupation function of the $\hat f$-fermion $n_f(\omega)$ by $h_f(\omega) = 1- 2 n_f(\omega)$; $\rho_f (\omega)$ is the spectral function of $\hat f$-fermion $\rho_f (\omega) = - 1/(2\pi) \Im \left[ G^>_f (\omega) - G^<_f (\omega)\right]$. Similarly, the Majorana Green's function is defined as:
\begin{equation}
G_\eta^>(\omega) = - i \pi \rho_\eta (\omega).
\end{equation}
where $\rho_\eta (\omega)$ is the spectral function of $\hat \eta$-fermion, $\rho_\eta (\omega) = - 1/(2\pi) \Im \left[ G^>_\eta(\omega) - G^<_\eta(\omega)\right]$.
{ Note that the anti-commutation relations $\lbrace f,f^\dagger\rbrace=1$ and $\lbrace \eta,\eta^\dagger\rbrace=2\eta^2 =2$ imply that $\int d\omega~\rho_f(\omega)=1$ and $\int d\omega~\rho_\eta(\omega)=2$.}

{ Our diagrammatic approach starts from the {\it bare} Hamiltonian $H=\omega_z f^\dagger f$, which is equivalent to
\begin{eqnarray}
\rho_f(\omega) &=& \delta(\omega-\omega_z),\; h_f(\omega) = 1 - 2 n_f(0), \\ \nonumber
\rho_\eta(\omega) &=& \delta(\omega).
\end{eqnarray}
{Here $n_f(0)$ can be understood as the occupation of fermions before coupling to the bath. It will not be important for the subsequent analysis.}

\paragraph{Self-energy corrections. --}
\begin{figure}[t]
\begin{centering}
\includegraphics[width=0.8\linewidth]{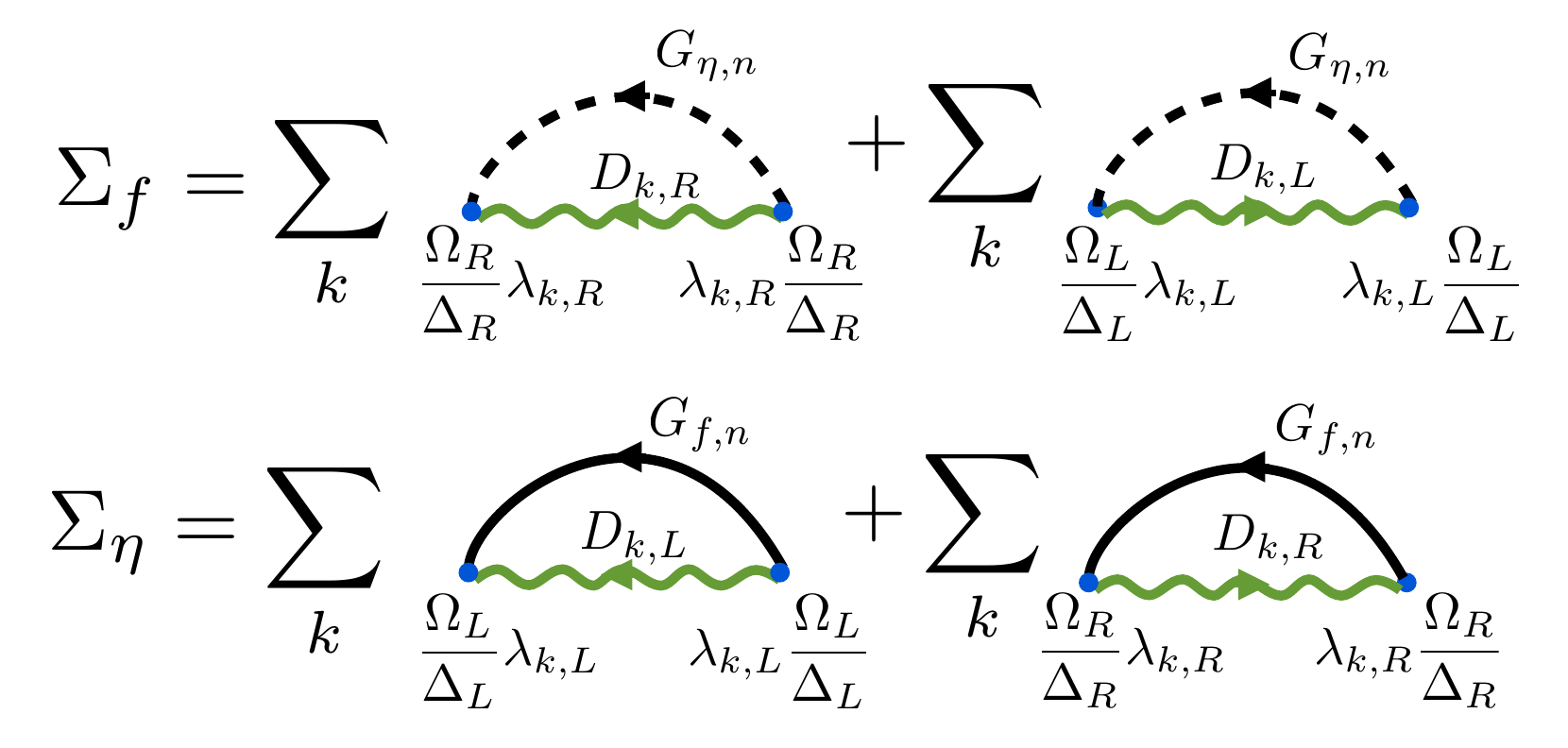}
\par\end{centering}
\protect\protect\caption{\label{fig:SE-2-3} Self-energies for the $\hat f$- and $\hat \eta$- fermions, $\Sigma_f$ and $\Sigma_\eta$. Solid lines corresponds to the Dirac fermion Green's function and dashed lines to the Majorana fermion Green's functions; wiggly lines represent the Green's functions of the bosonic bath.}
\end{figure}

We now calculate the impact of the dissipative bath coupled to the atoms. We consider the correction to the Green's function of the $\hat f$-fermion and $\hat \eta$-Majorana particle. The self-energy of the $\hat f$-fermion due to the interaction with the bath is given by the following expression (see the detailed derivation in Appendix~\ref{sec:app1})
\begin{eqnarray}
\Sigma_{f}^{R}(\omega)-\Sigma_{f}^{A}(\omega)  &=& - i (\gamma_{\uparrow}+\gamma_{\downarrow}) \\ \nonumber
\Sigma_{f}^{K}\left(\omega\right) &=& - i (\gamma_{\downarrow} - \gamma_{\uparrow})
\end{eqnarray}
In the steady state, the ratio between the self-energies on the Keldysh contour defines the function $h_f(\omega)$
\begin{eqnarray}
\label{eq:hf}
h_f(\omega)=\frac{\Sigma_{f}^{K}\left(\omega\right)}{\Sigma_{f}^{R}(\omega) -\Sigma_{f}^{A}(\omega) } = \frac{\gamma_{\downarrow} - \gamma_{\uparrow}}{\gamma_{\uparrow}+\gamma_{\downarrow}}
\end{eqnarray}

The polarization of the system is given by the equal time greater Green's function:
\begin{eqnarray} \label{eq:sz}
s_z(t) &\equiv& \frac{1}{2}\left( 2 f^\dag(t) f(t) -1 \right) =  \frac{1}{2}\left(  -2 i G^<_f(t,t) - 1 \right) \\  \nonumber
         &=& \frac{1}{2}\left(  -2 i \int \frac{d \omega}{2\pi} G^<_f(\omega) - 1 \right)
\end{eqnarray}
{By substituting the Greens' function $G^<_f(\omega)$ given by Eq.~\eqref{eq:GF_def} with $h_f$ defined by Eq.~\eqref{eq:hf} and taking the integral, we obtain}
\begin{eqnarray} \label{eq:spin_exp}
s_z=  -\frac{1}{2} \frac{\gamma_{\downarrow}-\gamma_{\uparrow}}{\gamma_{\uparrow}+\gamma_{\downarrow}}
\end{eqnarray}
Eq.~\eqref{eq:spin_exp} has a simple interpretation in terms of the spin model; { by definition (see Eq.~\ref{eq:spinrepr}) $s_z\equiv \langle S^z\rangle$ and the spin magnetization in the stationary state depends only on the ratio between the effective rate of the two dissipation channels.}

The self-energy of the $\hat \eta$-fermion due to the interaction with the bath is given by the following expressions (see Appendix~\ref{sec:app1}):
\begin{eqnarray}
&&\Sigma_{\eta}^{K}\left(\ve\right) = -2 i (\gamma_{\uparrow}+\gamma_{\downarrow}) \left( h_f (\omega) - \frac{\gamma_{\uparrow}-\gamma_{\downarrow}}{\gamma_{\uparrow}+\gamma_{\downarrow}} \right)\\ \nonumber
&& \Sigma_{\eta}^{R}(\omega) -  \Sigma_{\eta}^{A}(\omega) =  -2 i (\gamma_{\uparrow}+\gamma_{\downarrow})\left( 1 - h_f(\omega) \frac{\gamma_{\downarrow}-\gamma_{\uparrow}}{\gamma_{\uparrow}+\gamma_{\downarrow}} \right)
\end{eqnarray}
Substituting $h_f(\omega)$ given by the Eq.~\eqref{eq:hf}, we obtain
\begin{eqnarray}
\Sigma_{\eta}^{K}\left(\omega\right)  &=& 0  \\ \nonumber
\Sigma_{\eta}^{R}(\omega) -  \Sigma_{\eta}^{A}(\omega)  & = & - 2 i (\gamma_{\uparrow}+\gamma_{\downarrow})\left( 1- \left(\frac{\gamma_{\downarrow} - \gamma_{\uparrow}}{\gamma_{\uparrow} + \gamma_{\downarrow}} \right)^2\right)
\end{eqnarray}
Note, that when the bath does not have any coherence between its left and right part, it does not induce any anomalous terms
in the $\hat f$-fermion Green's functions, i.e. $\langle \hat f \hat f\rangle = \langle \hat f \hat \eta \rangle =0$.

 \subsection{Spin-spin correlation functions.}
\label{sub:spin_prop}

We now show how to use the Majorana fermion representation to compute the correlation functions of spin operators.
This calculation involves two distinct methods, depending on whether the expectation values of the spin operator under consideration is zero or non-zero. In the former case, the spin-spin correlation function can be expressed as a single Green's function, while in the latter case, the convolution of two Green's functions is required. This distinction was not fully appreciated in the earlier literature~\cite{Shnirman2003,mao2003spin,Biswas2011,Schad2015}.

In the Majorana fermion language, spin-spin correlations correspond to four-point correlation functions (see Eq.~\eqref{eq:spinrepr2}):
\begin{align}\label{eq:spin_spin_fermions}
\av{S^\alpha (t) S^\beta (t')} &=  \frac{(-1)^{n_\alpha + n_\beta}}{4}\av{ \hat \eta_\alpha(t)  \hat \tau_x(t) \hat \eta_\beta(t')  \hat \tau_x(t')}.
\end{align}
where $n_x = 1$, $n_y = 2$, and $n_z = 3$. For the sake of concreteness, we consider two spin operators, $\hat S^x$ and $\hat S^z$, whose expectation values respectively equal zero and non-zero.

{For the former operator,  $\av{S^x (t)} =\frac{1}{2}\av{\hat \eta_x (t) \hat\tau_x (t) } = 0$.} This implies that $\hat \tau$ and $\hat \eta_x$ fermions are uncorrelated and one can factorize their correlations. This allows us to break down the four operator average into the product of two operator averages.
\begin{eqnarray} \label{eq:spin_prop-corr}
\av{S^x (t) S^x (t')} &=& \frac{1}{4} \av{\tau_x(t)\tau_x(t')  } \\ \nonumber &&
\times\av{ \left( \hat f (t) +\hat f^\dag (t) \right)  \left( \hat f (t') + \hat f^\dag (t') \right)}.
\end{eqnarray}
{ Because $\tau_x$ commutes with the Hamiltonian (\ref{eq:sp12-ham}), it is invariant in time and $\langle\hat\tau_x(t)\hat\tau_x(t')\rangle =\langle \hat\tau_x^2\rangle =1$.} Thus, in the frequency domain, the correlation function can be represented solely by the $f$-fermion Green's function:
\begin{eqnarray} \label{eq:sxsx_gf}
\fourier{\av{S^x (t) S^x (t')}} &=& \frac{i}{4} G_f^>(\omega) -  \frac{i}{4} G_f^<(-\omega)
\end{eqnarray}
Substituting the expression for the lesser and greater Green's functions we obtain (see the detailed derivation in Appendix~\ref{sec:app2})
\begin{multline} \label{eq:sxsx_final}
\fourier{\av{S^x (t)S^x (t')}} =
 \frac{\gammad}{2}\frac{1}{\left( \omega-\omega_z\right)^2 + \left( \gammap\right)^2} +\\
\frac{\gammau}{2}\frac{1}{\left( \omega+\omega_z\right)^2 + \left( \gammap\right)^2}
\end{multline}

In the case of correlation function $ \av{S^z (t) S^z (t')}$, the expectation value of the spin is finite, $\av{\hat S^z} \neq 0$, and given by Eq.~\eqref{eq:spin_exp}. Thus, in this case, we are not allowed to decompose the four-fermion Green's function in the same way as in Eq.~\eqref{eq:spin_prop-corr}. To circumvent the difficulty of accounting for correlation between the $\hat \tau_x$ and $\hat \eta$ fermions we express the spin operator in terms of $f$ fermions, as $S^z = \hat f^\dagger \hat f - \frac{1}{2}$. Since there are no vertex corrections of second order in the coupling parameter $\lambda$, we can express the spin-spin correlation function as the product of two Green's functions.
\begin{multline}
   \av{S^z (t) S^z (t')} =  \av{S^z(t)}\av{S^z(t')}  \\ +G_f^<(t',t) G_f^>(t,t')
\end{multline}
In the stationary state we calculate the Fourier transform of the spin-spin correlation function, substitute the expressions for $G^>_f(\omega)$ and $G^<_f(\omega)$, and convolve two Green's function to obtain (see the detailed derivation in Appendix~\ref{sec:app2})
\begin{multline} \label{eq:szsz_final}
\fourier{   \av{S^z (t)S^z (t')}} = 2\pi \av{S^z}^2\delta(\omega) \\
   + \left(\frac{1}{4} - \av{S^z}^2\right) \frac{4 (\gamma_\uparrow + \gamma_\downarrow) }{\omega^2+4 (\gamma_\uparrow + \gamma_\downarrow)^2}
\end{multline}
These results are in agreement with the Lindblad approach analysis we provide in Sec.~\ref{sec:Lindblad}.

\section{Driven-dissipative 4-level scheme}
\label{sec:dis}

\subsection{The Model}
\label{sec:model}

In this section, we investigate a system consisting of a multi-level atom, coupled to a dissipative environment and driven externally by  laser fields. Specifically, we consider an atom with an internal structure represented by four states with energies $\ve_n$. Fig.~\ref{fig:Sketch_Spin}(b) shows the sketch of the system. Two pairs of states --  $\ket 1$, $\ket 3$ and $\ket 2$, $\ket 4$, --  are coupled using a coherent drive with frequencies $\omega_{R}$ and $\omega_{L}$ and matrix elements $\Omega_{R}$ and $\Omega_{L}$. In addition to the coherent drive, this 4-level system is coupled to incoherent bosonic baths that describe the decay of the states $\ket 3$ and $\ket 4$ to the states $\ket 1$ and $\ket 2$ respectively. The resulting scheme is often referred to as double-$\Lambda$ scheme, or W scheme, and was used by Ref.~\cite{Dimer2007} to offer a possible realization of the Dicke phase transition. The model was recently realized in a cavity QED experiment by Ref.~\cite{Zhiqiang2017}. The same scheme was used by Ref.~\cite{dallaotter} as a proposal to realize a spin-squeezed state: see their Supplementary Materials for two specific physical realizations using $^{87}$Rb atoms.

Our goal is to demonstrate how to treat coherent and dissipative processes on equal footing using non-equilibrium diagrammatic methods ~\cite{Keldysh1965}.
{ As a practical application of our method, we show how to use diagrammatic techniques to map this multi-level system to an effective Hamiltonian of a two-level system with dissipation.} At equilibrium, this mapping can be justified when the temperature is much smaller than the energy separation between ground and excited states. In the present non-equilibrium case, the temperature is not well defined. Nevertheless, the excited states can be integrated out if one assumes that (i) the atoms are initially prepared in the ground states $\ket 1$ and $\ket 2$, and (ii) the driving fields are far detuned from the resonances to states $\ket 3$ and $\ket 4$. Using the aforementioned conditions, we will demonstrate that we can integrate out virtually occupied degrees of freedom, and the rest of the system can be mapped to an effective two-level system with dissipation.

The Hamiltonian of the system can be written as { the} sum of Hamiltonians corresponding to all processes under consideration, $\hat H_{\rm ab}(t) = \hat{H}_{{\rm a}}(t)+ \hat{H}_{{\rm b}}  + \hat{H}_{\rm{ab, int}}$. Here $\hat{H}_{{\rm a}}(t)$ stands for the Hamiltonian of the atom
\begin{eqnarray}
\label{eq:H1}
\hat{H}_{{\rm a}} (t) & = & \sum_{n=1}^{4} \ve_{n} \hat c_n^\dag \hat c_n\\ \nonumber
&+&\left( \Omega_{L} e^{i\omega_{L}t} \hat c_1^\dag \hat c_4+ \Omega_{R}e^{i\omega_{R}t} \hat c_2^\dag \hat c_3  +\textrm{h.c.}\right),
\end{eqnarray}
where $\textrm{h.c.}$ is the Hermitian conjugate. This Hamiltonian includes processes induced by the external driving.

In Eq.~\eqref{eq:H1} we use the Schwinger-fermion representation of the states of the system. Here the operators $\hat c_n^\dag$ create { an electron in the state} $\ket n$. Note, that using these notations we should keep track of the number of electrons in the system which should be conserved and equal to one,
\begin{equation} \label{eq:constraint}
\sum_n^4 \hat c_n^\dag \hat c_n =1.
\end{equation}
In the following, we will derive an effective model and rewrite it using the Majorana fermion representation without requiring any constraints on the Hilbert space.

The coupling of the atomic system to the bosonic bath, $\hat{H}_{{\rm b}} $, is described though the interaction term $\hat{H}_{{\rm ab, int}}$, where
\begin{eqnarray}
	\label{eq:orig_haml}
\hat{H}_{{\rm b}} & = & \sum_{k,\sigma=\lbrace L,R\rbrace} \nu_{k} \hat d_{k,\sigma}^\dag \hat d_{k,\sigma},\\ \nonumber
\hat{H}_{{\rm ab, int}} & = &
            \sum_{k} \lambda_{L}(\hat d^\dag_{k,L}+\hat  d_{k,L}) \left(\hat  c_2^\dag \hat  c_4+ \hat  c_4^\dag \hat  c_2\right)\\ \nonumber
 & + & \sum_{k}\lambda_{R}(\hat  d^\dag_{k,R}+\hat  d_{k,R})  \left(\hat  c_1^\dag \hat  c_3+\hat  c_3^\dag \hat  c_1 \right).\nonumber
\end{eqnarray}
Here the bosonic operators $\hat d_{k,\sigma}^\dag$ and $\hat d_{k,\sigma}$ describe the processes of creation and annihilation of photons with frequency $\nu_{k}$ and polarization $\sigma = \lbrace L,R\rbrace$. The coupling between the atomic system and the photons is described { by} the interaction constants $\lambda_{L}$, $\lambda_{R}$, which are assumed to be small. For simplicity, we assume that the emitted photons have different polarization in the left and right channels and do not interfere with each other.

{ The operators $d_{k,\sigma}$ describe free EM modes with thermal occupation defined by temperature $T$. Their physical properties are then captured  by the correlation function (greater Green's function)} $D_{b,\sigma}^>(0,t) = \sum_k \lambda_{\sigma}^2\av{d_{k,\sigma} (t) d_{k,\sigma}^\dag (0) }$. This function can be written as a sum of its symmetric $S_\sigma(t)= S_\sigma(-t)$ and antisymmetric $A_\sigma(t)= -A_\sigma(-t)$ parts, $D_{b,\sigma}^>(0,t) = A_\sigma (t) + S_\sigma (t)$, { The components} $A_\sigma (t)$ and $S_\sigma (t)$ are associated with dissipation and fluctuations of the bosonic bath respectively. At thermal equilibrium, these functions are related by the fluctuation-dissipation theorem~\cite{kamenev2011field}
\begin{equation}
\frac{S_{\sigma} (\omega)}{A_{\sigma} (\omega)} = \coth\left( \frac{\omega}{2 T}\right)
\label{eq:FDT}
\end{equation}
where $A_{\sigma} (\omega)$ and $S_{\sigma} (\omega)$ are the Fourier transform of the corresponding time-dependent functions. Although the $d$ modes assumed to be at thermal equilibrium, the entire system is out-of-equilibrium due to the time-dependent driving term in Eq.~\eqref{eq:H1}.

\subsection{Rotating Wave Approximation}
\label{sec:RWA}

In the laboratory frame, the Hamiltonian describing the driven-dissipative system, $\hat H_{\rm{ab}}(t)$, is explicitly time-dependent due to the coherent driving. { To obtain an effective time-independent description of the problem, we  now move to a frame that rotates at the frequency $\omega_{\rm{dr}} = \frac{1}{2}\left(\omega_R+\omega_L \right)$.}
Mathematically, the transition to the rotating frame is performed by the transformation
\begin{eqnarray}
\hat c_1  &\to & \hat c_1, \;\; \hat c_2 \to \hat c_2 e^{-i \frac{\omega_{L}-\omega_{R}}{2} t},\;\; \\ \nonumber
\hat c_3  &\to & \hat c_3 e^{-i \omega_\text{dr}t},\;\; \hat c_4 \to \hat c_4 e^{-i\omega_{L}t},\;\;\\ \nonumber
\hat d_{k,\sigma} &\to & \hat d_{k,\sigma}  e^{-i\omega_{dr}t}
\end{eqnarray}
Under this transformation, the Heisenberg equation of motion of the new variables is determined by the following Hamiltonian,  $\hat H_{\rm ab}^\prime = \hat{H}_{{\rm a}}^\prime+ \hat{H}_{{\rm b}}^\prime  + \hat{H}_{\rm{ab, int}}^\prime$,
\begin{eqnarray}
	\label{eq:rwa_haml}
\hat{H}_{{\rm a}}^\prime & = & \sum_{n=1}^{4} \Delta_{n}\hat c_n^\dag \hat  c_n\\ \nonumber
&+&
\Omega_{L} \left( \hat c_1^\dag \hat  c_4+ \hat  c_4^\dag \hat  c_1 \right) +
\Omega_{R}\left( \hat  c_2^\dag \hat  c_3 + \hat  c_3^\dag \hat c_2\right)  \\ \nonumber
\hat{H}_{{\rm b}}^\prime & = & \sum_{\sigma,k} \nu_{k,\sigma}' \hat d_{k,\sigma}^\dag \hat d_{k,\sigma}\\ \nonumber
\hat{H}_{{\rm ab, int}}^\prime & = &
			\sum_{k} \lambda_{L}\left(\hat  d^\dag_{k,L} \hat  c_2^\dag \hat  c_4+\hat  d_{k,L} \hat c_4^\dag \hat  c_2\right)\\ \nonumber
 & + & \sum_{k}\lambda_{R}\left(\hat  d^\dag_{k,R} \hat  c_1^\dag \hat  c_3+\hat  d_{k,R}\hat  c_3^\dag \hat c_1 \right).\nonumber
\end{eqnarray}
where $\Delta_{1} = \varepsilon_1$, $\Delta_{2}=\ve_{2}-{(\omega_{L}-\omega_{R})}/{2}$, $\Delta_{3}=\ve_{3}-{(\omega_{L}+\omega_{R})}/{2}$, $\Delta_{4}=\ve_{4}-\omega_{L}$.
Here we neglected the counter-rotating terms of the light-matter interaction $H_{int}$ which oscillate { at the optical} frequency $\omega_{dr}$ (see Appendix ~\ref{sec:app4} for more details).

In the rotating frame, the eigenfrequencies of the bosonic baths are shifted from the original ones by
$\nu_{k,\sigma}' =\nu_{k,\sigma} - \omega_\text{dr}$. { This results in a modified fluctuation-dissipation relation: Eq.~\eqref{eq:FDT} becomes}
\begin{equation}
\frac{S_{R}^\prime (\omega)}{A_{R}^\prime (\omega)} = \frac{S_{R} (\omega + \omega_\text{dr} )}{A_{R} (\omega + \omega_\text{dr} )} = \coth\left( \frac{\omega + \omega_\text{dr}}{2 T}\right)\;.
\end{equation}
{ In quantum optical systems, the driving frequency $\omega_\text{dr}\sim 10^{15}$~Hz is the largest frequency in the system, and in particular it is much larger than the typical interaction scale $\omega \sim 10^3-10^9$~Hz and the temperature of the bath $T\approx 300$~K~$\sim 10^{12}$~Hz (for room temperature experiments). Under these two conditions one can safely approximate  $\coth\left( \omega_\text{dr}/2T\right) \approx 1$.} This approximation is equivalent to the common Born-Markov approximation used in the master equations' approach~\cite{scully1999quantum}.

{ Under this approximation, the following relations between the antisymmetrized and symmetrized parts of the correlation function of the two baths can be established
\begin{eqnarray}\label{eq:BATH}
A_\sigma^\prime(\omega) &=& S_\sigma^\prime(\omega) = A (\omega_\text{dr})
\end{eqnarray}
Let us stress that both $A_\sigma(\omega)$ and $S_\sigma(\omega)$ are symmetric with respect to $\omega\to -\omega$. This makes the Markovian bath different from zero-temperature ones,   where $S(\omega) = A(\omega) {\rm sign}(\omega)$ and their product is always anti-symmetric}.

\subsection{Two-level effective model: adiabatic elimination approach}
\label{sec:lowenegy}

We now assume that only two states of the atomic system are physically occupied. This allows us to derive an effective two-level model with dissipation.
We eliminate the virtually occupied states, $\ket 3$ and $\ket 4$, by using an elimination procedure based on the path-integral technique~\cite{brion2007adiabatic}. This step is equivalent to the common ``adiabatic elimination'' used in the context of Markovian master equations~\cite{cohen1992atom,scully1999quantum}. Specifically, we represent the system using Grassmann variables in the path-integral approach. The part of the action containing states  $\ket 3$ and $\ket 4$ is quadratic in the corresponding fermions $\hat c_3$ and $\hat c_4$ with a linear coupling to the other states. We use the Gaussian integral identity,
\begin{eqnarray}\label{eq:Beresin}
\int d\bar{c} \; dc \; e^{-i \int_C \bar{c} \; G_c^{-1} c + \bar c \; V + \bar V c} =
 \det G_c^{-1}e^{-i  \int_C \bar V G_{c} V }
\end{eqnarray}
where $c$ and $\bar c$ are Grassmann variables, $G_c(t,t')$ is the unperturbed (bare) Green's function corresponding to these variables, and $V(t)$ represents linear couplings to the rest of the other degrees of freedom of the system. The integral $\int_C $ represents the integration along the Keldysh contour.

With the help of Eq.~\eqref{eq:Beresin} we integrate the variables that correspond to states $\ket 3$ and $\ket 4$. If the integration is performed exactly, the problem becomes non-Hamiltonian due to retardation effects that come into play.
This complication can be avoided if the pumping drives are far detuned from the excited state.
In particular, the bare Green's function of the fermions  $\hat c_3$ and $\hat c_4$ in the rotating frame  reads
\begin{equation}\label{eq:Gfbare}
G_{c}^R(\omega) = \frac{1}{(\omega-\Delta_c)}
\end{equation}
where $\Delta_c$ is the detuning of the corresponding state. When the detuning $\Delta_3$($\Delta_4$) is larger than all relevant energies, $\omega \ll \Delta $, the Green's function~\eqref{eq:Gfbare} can be approximated with an expression local in time:
\begin{equation}\label{eq:NoRetardation}
G_c(\omega) \approx \frac{1}{\Delta_c}  \rightarrow \; G_{c}(t,t')  \approx \frac{1}{\Delta_c} \delta(t-t').
\end{equation}
In this approximation, there are no retardation effects and the problem remains Hamiltonian (see Appendix \ref{sec:app5} for more details). This derivation also shows how to extend our analysis to the non-Markovian case by taking into account higher-order terms in $\omega$. This is the Keldysh analogue of the common derivation using master equations, where the non-trivial Nakajima-Zwanzig formalism is required to take into account higher order terms~\cite{breuer2002theory}.

Using expression~\eqref{eq:Beresin} and approximation~\eqref{eq:NoRetardation} we obtain the new effective Hamiltonian.
We recast it in the following form, $\hat{\mathcal{H}}_\text{ab} = \hat{\mathcal{H}}_{{\rm a}}  + \hat{\mathcal{H}}_{{\rm b}} +\hat{\mathcal{H}}_{{\rm ab,int}}$
\begin{eqnarray} \label{eq:Ham3}
\hat{\mathcal{H}}_{{\rm a}} & = & \omega_z c_2^\dag c_2\\ \nonumber
\hat{\mathcal{H}}_{{\rm b}} & = & \sum_{\sigma,k}{\omega}_{k,\sigma} d_{k,\sigma}^\dag d_{k,\sigma}\\ \nonumber
\hat{\mathcal{H}}_{{\rm ab,int}} & = &
        \sum_{k}   \frac{\Omega_{L}}{\Delta_L} \lambda_{L}\left(d^\dag_{k,L} c_2^\dag c_1+d_{k,L} c_1^\dag c_2\right)\\ \nonumber
 & + & \sum_{k}\frac{\Omega_{R}}{\Delta_R}\lambda_{R}\left(d^\dag_{k,R} c_1^\dag c_2+ d_{k,R} c_2^\dag c_1 \right)\nonumber
\end{eqnarray}
where we introduced $\omega_z = \Delta_2 -\Delta_1 + \frac{\Omega_{R}^2}{\Delta_R} - \frac{\Omega_{L}^2}{\Delta_L}$.  After the adiabatic elimination, the term related to the bosonic bath, $\hat{\mathcal{H}}_{{\rm b}}$, remains unchanged.

Note that the constraint in Eq.~\eqref{eq:constraint} should be fulfilled. After the elimination of the states $\ket 3$ and $\ket 4$, this constraint reads
\begin{equation}\label{eq:constr2}
c_1^\dagger c_1 + c_2^\dagger c_2= 1.
\end{equation}
It  shrinks our physical Hilbert space to the two states $|n_1=1, n_2=0\rangle$ and $|n_1=0, n_2=1\rangle$.
The Hamiltonian \eqref{eq:Ham3} and the constraint \eqref{eq:constr2} is equivalent to a spin-$1/2$ system coupled to Markovian baths and can be analyzed with the tools developed in Sec.~\ref{sec:sp}.

\section{Ensemble of atoms interacting with a single-mode optical cavity}
\label{sec:phasetr}

\subsection{Effective model}
\label{sub:eff_model}

In the previous Secs.~\ref{sec:sp} and \ref{sec:dis}, we discussed the diagrammatic description of a single multi-level system coupled to a dissipative bath driven by an external field.  We demonstrated how to describe properties of the system using the fermionic representation of the system. This formalism is readily applicable to the description of driven-dissipative ensembles of atoms interacting with bosonic fields.  In particular, we provide an example of an open quantum system of $N$ four-level atoms interacting with a single optical mode cavity. We consider the dissipation of the cavity, as well as the dissipation of each four-level atom. Fig.~\ref{fig:Sketch_Spin}(c) shows a sketch of the system. The scheme was proposed as a realization of the generalized Dicke model in Ref.\cite{Dimer2007,Bhaseen2012}.

We describe the system with the Hamiltonian consisting of three parts, ${\hat{H}} = {\hat{H}}_\text{a} + {\hat{H}}_\text{c} + {\hat{H}}_\text{int}$, where the atomic part of the system, the optical cavity, and the interaction are characterized by ${\hat H}_\text{a}$, ${\hat{H}}_\text{c}$, and  ${\hat{H}}_\text{int}$ correspondingly. The atomic part of the system is a sum over independent single atom Hamiltonians interacting with a dissipative bath ${\hat{H}}_\text{a} = \sum_n^N \hat{H}_{a,0}$, where $\hat H_{a,0}$ is given by Eq.~\eqref{eq:H1}. The optical mode is described as a single Harmonic oscillator with frequency $\omega_{0}$, coupled to a dissipative bath, described by the Hamiltonian \begin{align}\hat H_\text{c} = \omega_0 a^\dag a{+\sum_k\omega^b_kb^\yd_kb^\nd_k+\sum_k\kappa_k\left(b^\yd_ka+b^\nd_ka^\yd\right)}.\end{align}
Here { the} operators $\hat a^\dag$ and $\hat a$ represent the creation and annihilation of the cavity photons, $b^\yd_k$ and $b^\nd_k$ represent the creation and annihilation operators for the cavity bath. The coupling $\kappa_k$ is defined such that $\sum\nolimits_k\frac{\kappa_k^2}8B^>_{k}(\omega)=-i\kappa$, where $B^>(k)$ is the larger Green's function of the bath and $\kappa$ is the cavity decay rate, in analogy to Eq.~\eqref{eq:bath_properties1}.

The interaction of the atomic system with the cavity mode is analogous to the atom-bath Hamiltonian~\eqref{eq:H1}:
\begin{eqnarray}\label{eq:HintDicke_lab}
\hat{H}_{{\rm int}} & = &
	    \sum_{n=1}^N {\eta_L}\left(a^\dag c_{n,2}^\dag c^{\nd}_{n,4}+a c_{n,4}^\dag c^{\nd}_{n,2}\right)\\ \nonumber
	& + & \sum_{n=1}^N {\eta_R}\left(a^\dag c_{n,1}^\dag c^{\nd}_{n,3}+a c_{n,3}^\dag c^{\nd}_{n,1} \right)\nonumber,
\end{eqnarray}
 where $\eta_{L}$ and $\eta_{R}$ are the atomic couplings between the high-energy and low-energy states. { The coupling between the atoms and the cavity field (\ref{eq:HintDicke_lab}) is similar to the coupling to dissipative channels in Eq. (\ref{eq:Ham3}), but there
is an important distinction.
{ The
difference concerns the coherence between the right and left part of the system. The dissipative modes typically do not show any coherence between the photons spontaneously emitted in the right and left atomic dissipation channels. In contrast, here we assume that the cavity photons emitted in the right and left channels are coherent.}

We now proceed in full analogy with the derivation in Sec.~\ref{sec:dis}. First, we rewrite the Hamiltonian in the rotating frame and use the rotating wave approximation. Second, we eliminate the virtual states and derive an effective Hamiltonian. Lastly, we rewrite this Hamiltonian using the mixed Majorana fermions representation.
After these transformations, the atomic Hamiltonian $H_a$ becomes equivalent to a sum over terms of the form of Eq.~(\ref{eq:Hamilt_diss}), the cavity frequency is shifted to
\begin{equation}\omega_\text{c} = \omega_{0}-\omega_\text{dr}+\frac{\eta_\text{R}^2}{ \Delta_\text{R}} -\frac{\eta_\text{L}^2}{ \Delta_\text{L}}.\end{equation}
and the interaction term becomes
\begin{eqnarray} \label{eq:Dicke_int}
\hat{\mathcal{H}}_{{\rm int}} & = &
	\sum_{n}^N {\lambda} \left( a^\dag \eta_n f_{n} + a f_n^\dag\eta_n \right)\\&+& \nonumber
	\sum_{n}^N {\lambda'} \left( a \eta_n f_n + a^\dag f_n^\dag \eta_n\right)
\end{eqnarray}
where the effective couplings are ${\lambda}=\Omega_R \eta_R \Delta_R^{-1}$ and
${\lambda'}=\Omega_L \eta_L \Delta_L^{-1}$ respectively for the rotating and counter-rotating terms. This Hamiltonian can be written more compactly using the Nambu notation. If we represent the creation and annihilation operators of fermions and bosons with the vectors $\hat{\textbf{f}}^T = \left( \hat f , \hat f^\dag\right) $ and $\hat{\textbf{a}}^T = \left( \hat a , \hat a^\dag\right)$ correspondingly, we can rewrite Eq.~\eqref{eq:Dicke_int} as
\begin{eqnarray}\label{eq:Nambu}
\hat H_\text{int}=  \hat{\textbf{a}}^\dag \Lambda\hat{\textbf{f}} \; \hat \eta, \;\;\;\; \Lambda = \left(\begin{array}{cc} -{\lambda} & {\lambda'} \\ -{\lambda'} & {\lambda}\end{array}\right),
\end{eqnarray}
where $\Lambda$ is the interaction matrix.

To help the comparison with related works, we point out that the resulting Hamiltonian can be written in the spin notation as
\begin{align}
        \label{eq:gDM}
	\hat H &= \hat H_\text{a} + \hat {H}_\text{c}
	+ \hat H_\text{int},
\end{align}
with
\begin{eqnarray}
        \label{eq:nonumber}
\hat H_\text{a} &=& \sum_n^N \omega_0 \hat \sigma^z_n  + \sum_{\sigma,k}\nu'_{k,\sigma} d_{k,\sigma}^\dag d_{k,\sigma}\nonumber\\ \nonumber
 & + & 2\sum_{n}^N\sum_{k}   \frac{\Omega_{L}}{\Delta_L} \lambda_{L}\left(d^\dag_{k,L} S^+_n+d_{k,L} S^-_n\right)\\
 & + & 2 \sum_{n}^N\sum_{k}    \frac{\Omega_{R}}{\Delta_R}\lambda_{R}\left(d^\dag_{k,R}  S^-_n + d_{k,R}  S^+_n \right)\\
\hat H_\text{c} &=& \omega_0 a^\dag a{+\sum_k\nu'_kb^\yd_kb^\nd_k+\sum_k\kappa_k\left(b^\yd_ka+b^\nd_ka^\yd\right)}\nonumber\\
\nonumber
\hat{{H}}_{{\rm int}} & = &
\sum_{n}^N {2\lambda} \left( a^\dag S^-_n + a S^+_n \right)+
\sum_{n}^N {2\lambda'} \left( a S^-_n + a^\dag S^+_n \right).
\end{eqnarray}}
This Hamiltonian is known as the generalized open Dicke model~\cite{Dimer2007,Bhaseen2012} and has two important limiting cases: (i) ${\lambda}={\lambda'}$ is the limit of the Dicke model~\cite{Dicke,HeppLieb} (ii) in the limit ${\lambda'}=0$ the model is equivalent to the Tavis-Cumming model~\cite{TavisCummings}, the many-body version of the Jaynes-Cummings model~\cite{JaynesCummings}.

\begin{figure}[t]
\begin{centering}
\includegraphics[width=1\linewidth]{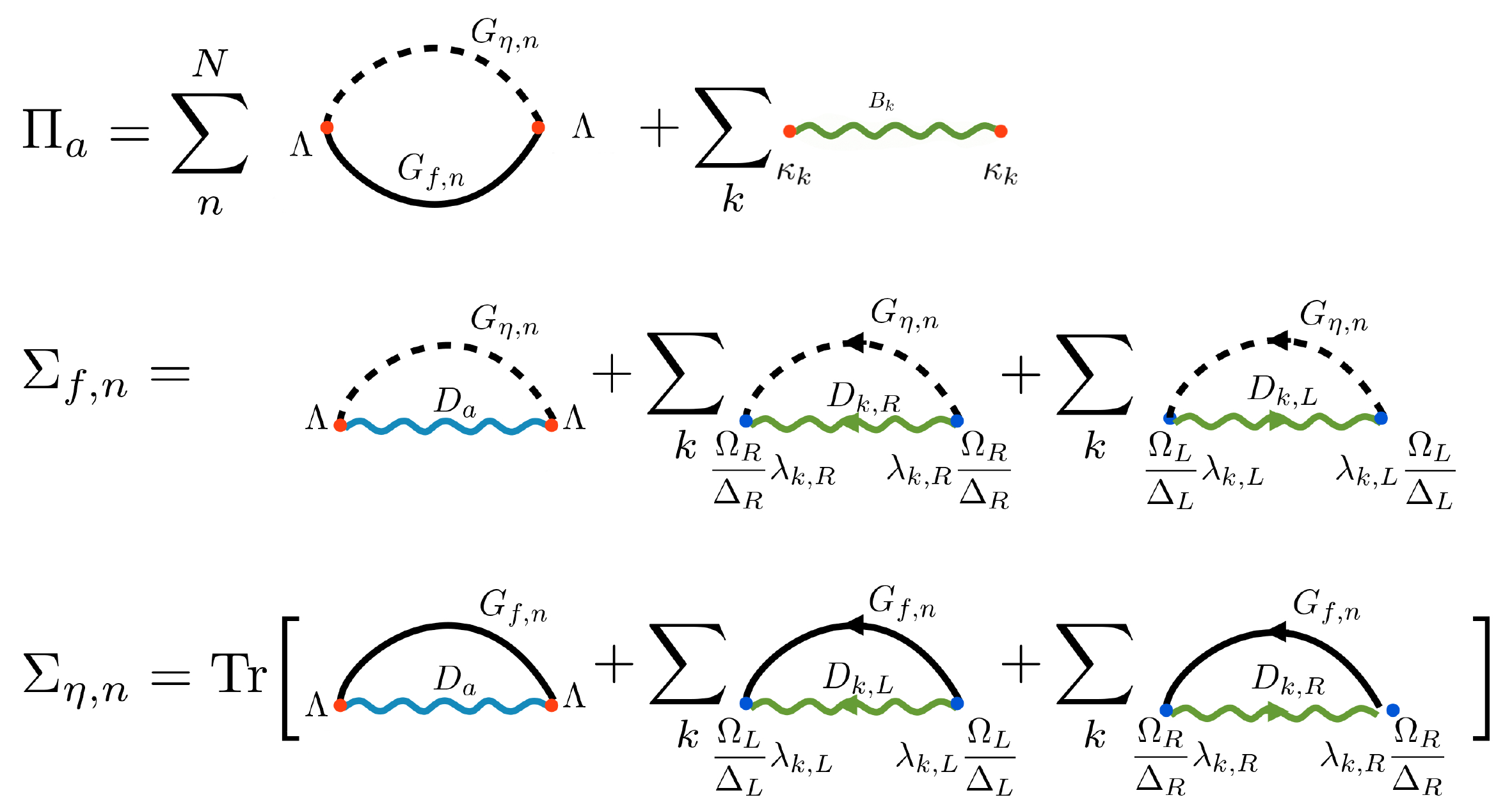}
\par\end{centering}
\protect\protect\caption{\label{fig:Dicke_diagrams} Self-energy contributions to the Green's functions of the cavity photons, $\Pi_a$, the Majorana fermions, $\Sigma_{\eta,n}$, and the Dirac fermions, $\Sigma_{f,n}$. }
\end{figure}

\subsection{Diagrammatic approach}
\label{sub:diag_dicke}

We are interested in the description of the { steady-state phase diagram of the system}. In particular, in the limit of the Dicke model, ${\lambda}={\lambda'}$, there is a phase transition between a normal and a superradiant phase~
\cite{sarang10,nagy11,oeztop11,strack11,bastidas12,Bhaseen2012,DallaTorre2013,buchhold2013dicke,zhang2013finite,keeling2014fermionic,
Buijsman2017,Gelhausen2017,Kirton2017,Safavi-Naini2017,Liu2017}.
The transition takes place when the interaction of the cavity mode with the atomic system softens the cavity mode.
At the transition the system becomes unstable with respect to the normal phase, thus the cavity gains a macroscopic occupation.

In Fig.~\ref{fig:Dicke_diagrams}, we show the self-energy contributions to the cavity photons, and the Majorana and Dirac fermion Green's functions ($\Pi_a$, $\Sigma_{\eta,n}$ and $\Sigma_{f,n}$). These contributions come from the various processes described by the Hamiltonian $\hat{H}$, which include{s} the cavity-atom coupling \eqref{eq:Dicke_int}, as well as the coupling of the atoms and of the cavity to incoherent bosonic baths. Note that the self-energies of the Dirac fermions $\Sigma_{f,n}$ and the cavity photon $\Pi_{a}$ are matrices whose elements are calculated by the matrix multiplication of the interaction constant $\Lambda$ (Eq.~\eqref{eq:Nambu}) with the corresponding Green's function. In contrast, the self-energy of the Majorana fermion is a scalar; it is given by the trace of the corresponding self-energy matrix.

Different contributions to the self-energies are classified according to their scaling with the
 coupling strength $\Lambda$ and the number of atoms $N$
(see Fig.~\ref{fig:Dicke_diagrams}). In particular, the cavity photons are coupled to the atomic system by the generalized Dicke interaction term~\eqref{eq:Dicke_int}. The contribution from each atom is proportional to ${\lambda^2}$. Under realistic assumptions, $\lambda$ is very small and the self-energy contributions from a single atoms are negligible. However, by summing up the contributions from $N$ atoms, one obtains a self-energy proportional to ${N\Lambda^2}$. For $N\gg 1$, this collective contribution can have a significant effect on the cavity \cite{kockum2019ultrastrong}.

The Dirac fermion's self-energy has two types of contributions. The first contribution comes from the interaction with the cavity photon and is proportional to ${\lambda^2}$. The second contribution comes from the interaction of the atom with the incoherent bosonic bath and does not depend on the
 coupling strength%
. Thus, in the limit of
 small $\lambda$,
the first contribution can be neglected.

The scaling arguments presented above can be put on a solid theoretical ground by introducing the rescaled effective couplings ${g}=\sqrt{N}\lambda$. and ${g'}=\sqrt{N}\lambda'$. Using this notation, the interaction term in Eq.~(\ref{eq:nonumber}) reads
\begin{align}
	\label{eq:Hamil_with_g}
		H_{\rm int}=&\sum_n^N\frac{2g}{\sqrt{N}}\left(a^\yd S^-_n+a^\nd S^+_n\right)+\sum_n^N\frac{2g'}{\sqrt{N}}\left(a^\nd S^-_n+a^\yd S^+_n\right),
\end{align}
The model is then studied in the limit of $N\to\infty$, while keeping fixed $g$ and $g'$. This procedure allows one to perform a controlled resummation of a specific subset of diagrams, whose prefactor does not tend to zero in the limit of $N\to\infty$ \cite{DallaTorre2013,piazza2013bose,piazza2014quantum,lang2016critical,DallaTorre2016}. From a physical perspective, this ``large-$N$'' approximation is equivalent to neglecting the feedback of the cavity on the spins.

We now focus on the non-equilibrium steady state of the system in the long time limit. {The equations for the out-equilibrium dynamics of the system are provided in Appendix~\ref{sec:app3}. In practice, the calculation should be organized as follows. First, the correction to the Green's functions of the fermions due to the interaction with the dissipative bath should be calculated. Then those Green's functions are used for calculating the correction to the cavity photon Green's function.

In Sec.~\ref{sec:diagr}, we provided the calculation of the self-energies of the fermion interaction with the dissipative bath. We use those calculations as an initial point for calculating the self-energy of the cavity photons. We notice that in the lowest order
\begin{eqnarray}\label{eq:pi_gfgeta}
\left[ \Pi_{a}^{R}(\omega)\right]_{\alpha \alpha'} & = &  \sum_{n}^N \sum_{\beta \beta'} \Lambda_{\alpha \beta}^T \times\\ \nonumber &&
\int d\ve  \left[ G_{f,n}^K(\omega-\ve)\right]_{\beta \beta' } G_{\eta,n}^R(\ve) \Lambda_{\beta' \alpha'}
\end{eqnarray}
where the Greek indexes correspond to the matrix elements in the Nambu space and $\Lambda$ is defined above in Eq.~\eqref{eq:Nambu}. {In fact, we notice that the self-energy of the cavity photon can be interpreted in terms of spin-spin correlation functions~\cite{DallaTorre2016}. Indeed, using the connection between fermion and spin representations~\eqref{eq:spinrepr} we can rewrite the self-energy as
\begin{multline}\label{eq:polariz0}
 \left[ \Pi_{a}^{R}(\omega)\right]_{\alpha \alpha'}  =     8 \sum_{n}^N \sum_{\beta \beta'} \Lambda_{\alpha \beta}^T \times \\
	\left[\left(\begin{array}{cc} \fourier{\av{\left[\hat S^- (0), \hat S^+(t)\right]}} & 0 \\0 &   \fourier{\av{\left[\hat S^+ (0), \hat S^-(t)\right]}} \end{array}\right) \right]_{\beta \beta' }\Lambda_{\beta' \alpha'}
\end{multline}
where $\fourier{\av{\left[\hat S^- (0), \hat S^+(t)\right]}}$ is the spin response function at frequency $\omega$. As we showed previously in Sec.~\ref{sub:spin_prop}, the calculation of the spin-spin correlation functions can be simplified using the Majorana fermions representation. In particular, for the case of spin response function, we have $\fourier{\av{\left[\hat S^- (0), \hat S^+(t)\right]}} = G_f^K(\omega) / 4$.}

Substituting $ G_f^K(\omega)$ in Eq.~\eqref{eq:polariz0} we obtain the following expression for the self-energy of the cavity photons:
\begin{eqnarray}\label{eq:polariz}
\left[ \Pi_{a}^{R}(\omega)\right]_{\alpha \alpha'}  & = &    2  \sum_{n}^N \sum_{\beta \beta'} \Lambda_{\alpha \beta}^T \times \\ \nonumber &&\left[\left(\begin{array}{cc} \frac{s_z}{\omega-\omega_z + i \Gamma}& 0 \\0 &   \frac{s_z}{\omega+\omega_z + i\Gamma}\end{array}\right) \right]_{\beta \beta' }\Lambda_{\beta' \alpha'}
\end{eqnarray}
where $\Gamma=\gamma_\uparrow+\gamma_\downarrow$ is the relaxation rate of the $\av{\left[S_x(0), S_x(t)\right]}$ response function. Note that as in the calculation of the response functions, only the poles in the lower complex half plane contributes to Eq.~\eqref{eq:polariz}.

We calculate the Green's function of the cavity photons using the Schwinger-Dyson equation~\cite{Dyson1949,Schwinger1951}, $D_a^{-1} =  \Pi_a +  D_{a,0}^{-1}$ (where $D^{-1}_{a,0}$ is the inverse bare Green's function)
\begin{eqnarray} \label{eq:GF_photons}
\left[ \left(D_a^R(\omega)\right)^{-1}\right]_{\alpha \alpha'} && =  \left[ \Pi_{a}^{R}(\omega)\right]_{\alpha\alpha'} +  \\ \nonumber &&
\left[ \left(\begin{array}{cc} \omega - \omega_c  + i \kappa  &  0\\  0&  -\omega -\omega_c - i\kappa \end{array}\right)\right]_{\alpha\alpha'}.
\end{eqnarray}
Here the coefficient $\kappa$ includes all the cavity leaking processes.

\subsection{Superradiant transition}
\label{sub:phase_tr}

One of the key properties of the Dicke model is the presence of a phase transition between a normal phase and a superradiant phase, known as the superradiant transition of the Dicke model. (Not to be confused with Dicke superradiant which occurs in free space -- see also Ref.~\cite{kirton2018introduction} for an more details.) The phase transition manifests itself as a dynamical instability of the cavity and, hence, can be detected by considering its response function, namely the retarded Green's function $G^R_a$. At the superradiant phase transition one of the poles of the Green's function crosses the origin of the complex plane and acquires a positive imaginary value. At this point, the response becomes an exponentially growing function of time, indicating a dynamical instability. For a Gaussian theory, the poles of the retarded Green's function $G^R$ correspond to the zeros of the inverse Green's function $D^R$, leading to the following condition for the  superradiant transition: \begin{align} \lim_{\omega\rightarrow 0 }\det\left[D^R_a(\omega)\right] = 0.\end{align}

Substituting the Green's function~\eqref{eq:GF_photons} into this expression we obtain the following algebraic equation for the {superradiant} phase transition
\begin{eqnarray} \label{eq:phase_tr}
&& 4  s_z^2 \left(g^2 - (g')^2\right)^2  +\left( \kappa ^2 +\omega _0^2\right)\left(  \Gamma ^2+ \omega _z^2\right)+  \\ && \nonumber
4 \omega _0 \omega _z  s_z  \left( g^2 + \left(g'\right)^2\right)-4  \kappa  \Gamma  s_z\left( g^2 - (g')^2\right) = 0,
\end{eqnarray}
{where we introduced} the total decay rate {$\Gamma=\gamma_\uparrow+\gamma_\downarrow$, and we recall that }$s_z$ is given by Eq.~\eqref{eq:sz}. The latter two parameters are controlled by the incoherent bosonic bath {which} is coupled to the atomic system. Solving
Eq.~\eqref{eq:phase_tr} for $g$ and $g'$ with fixed parameters $\Gamma$, $s_z$, $\kappa$, $\omega_c$, and $\omega_z$ {provides {a} critical line of the superradiant transition}.

For a fixed $g/g'$ ratio, Eq.~\eqref{eq:phase_tr} predicts that the Dicke transition occurs at
\begin{eqnarray}\label{eq:phase_tr-g_crit}
g_\text{crit}^2 &&= \frac{(R-1) \left(\omega _c \omega _z+  R \kappa \Gamma \right)}{4 R^2 s_z } \times \\ \nonumber &&
\left( 1-\sqrt{1-\frac{R^2 \left(\omega _c^2+\kappa ^2\right) \left(\Gamma ^2+\omega _z^2\right)}{\left(\omega _c \omega _z+R \kappa  \Gamma \right){}^2}}\right)
\end{eqnarray}
where $R = (1-(g'/g)^2)/(1+(g'/g)^2)$.

In the limit $g\to g' $, $R \to 0$ and Eq.~\eqref{eq:phase_tr-g_crit} reproduces the expression for the critical coupling $g_\text{crit}^2 = - \frac{\left(\omega _c^2+\kappa ^2\right) \left(\Gamma ^2+\omega _z^2\right)}{8 \omega _c  \omega _z s_z}$ for the Dicke model with the cavity and spin dissipation processes~\cite{DallaTorre2016,Kirton2017,Gelhausen2017}. For $g'=0$ ($R = 1$), the critical coupling does not have real solutions. Thus, we recover the result of the Tavis-Cummings model, which does not have a transition in the presence of dissipation \cite{larson2017some}.

\begin{figure}[t]
\begin{centering}
\includegraphics[width=1\linewidth]{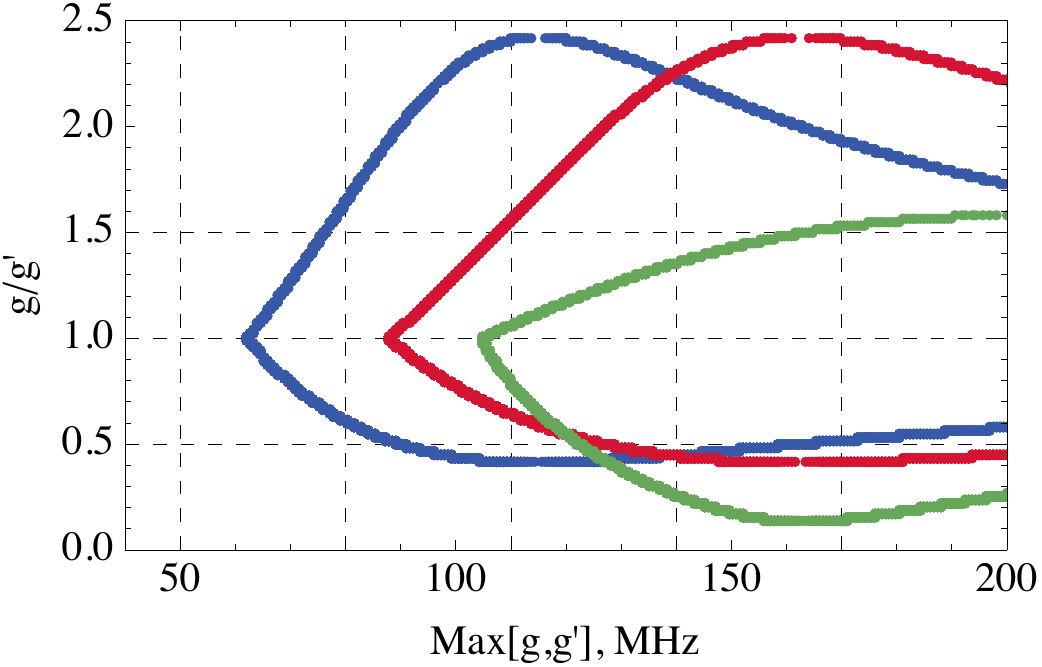}
\par\end{centering}
\caption{\label{fig:diag1} Critical line of the generalized Dicke model with dissipation. Different colors represent different parameters of the system. We considered the cavity mode with frequency $\omega_0=100$~kHz and dissipation $\kappa=100$~kHz, with the two-level splitting $\omega_z = 77.2$~kHz. The blue line corresponds to the case $\Gamma =0$, $s_z = -0.5$. The red line is $\Gamma =0$, $s_z = -0.25$, and the green line $\Gamma=50$~kHz, $s_z = -0.25$.}
\end{figure}

In Fig.~\ref{fig:diag1} we show the critical lines of the generalized Dicke model with dissipation for parameters relevant to the experiment of Ref.~\cite{Zhiqiang2017} (see the discussion in Sec.~\ref{sec:experiment}).
We show how the values of $g$ and $g'$ vary with the spin decoherence rate $\Gamma$ and polarization $s_z$. We specifically consider three limiting cases: (i) the limit of zero atomic dissipation $\Gamma =0$ and fully polarized initial state $s_z = -1/2$; (ii) zero atomic dissipation $\Gamma = 0$ and partially polarized initial state $s_z = -0.25$; and (iii) small atomic dissipation with a steady state polarization $s_z = -0.25$. The critical line for different polarization and dissipation rates shows a qualitatively similar behavior. The minimal critical coupling is achieved when $g=g'$, which corresponds to the case of the Dicke model. { The critical coupling increases when the ratio between the rotating and counter-rotating terms becomes either larger or smaller than 1}. Indeed, in both limits of $g/g'\gg 1$ and $g/g'\ll1$, the systems become equivalent to the Tavis-Cumming model and superradiance cannot be achieved. Furthermore, when initially the system is not in the fully polarized state, the critical line is shifted to the higher coupling strength. This effect is ultimately due to the fact that the spin response function $\fourier{\av{ \left[ S^{\mp}(0), S^{\pm}(t)\right] }} $ is proportional to  the polarization of the system, see e.g. Eq.~\eqref{eq:polariz}. Thus, it is natural to expect that partially polarized systems are less superradiant. Adding the dissipation makes the critical line less symmetric with respect to the $g=g'$ line. By its nature, atomic dissipation decreases the effect of the counter-rotating terms. Hence, larger coupling to the counter-rotating terms $g'$ is required to get to the superradiant phase.

\subsection{Stability diagram}
\label{sub:stability}

\begin{figure*}[t]
\begin{centering}
\includegraphics[width=2\columnwidth]{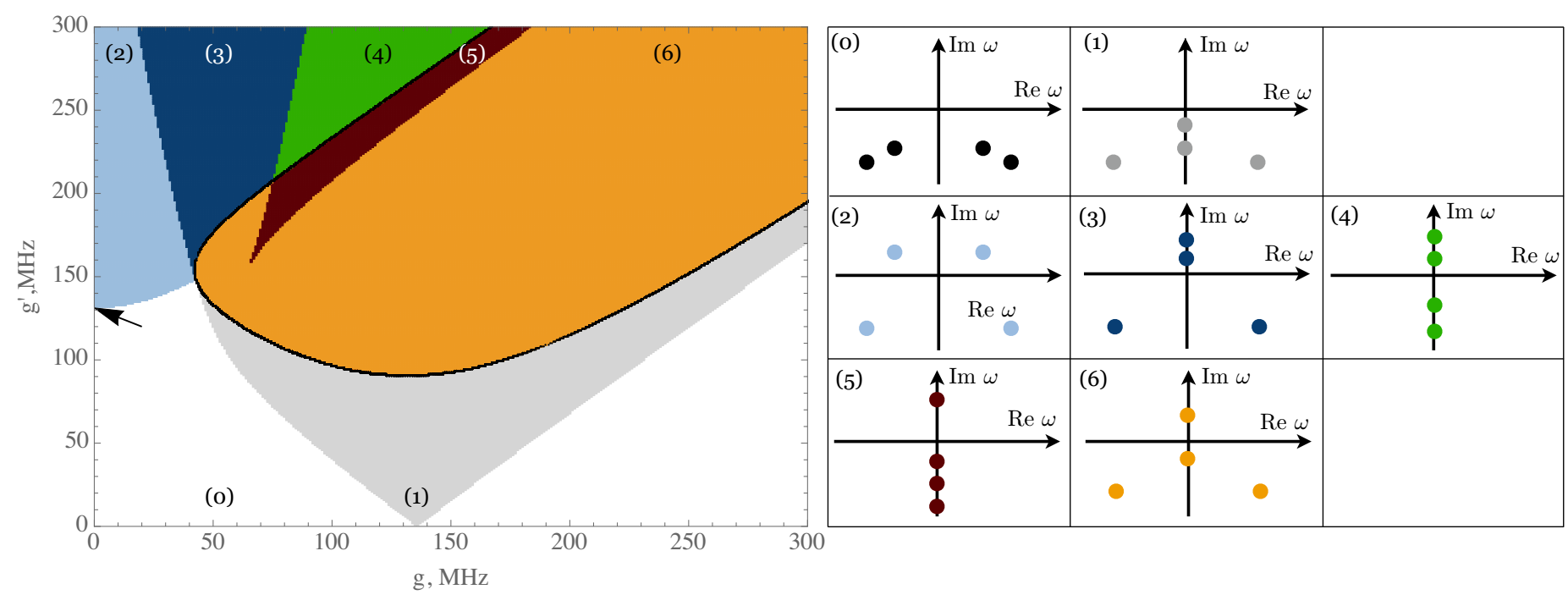}
\par\end{centering}
\caption{\label{fig:diag2} Left panel: Stability diagram of the generalized Dicke model with dissipation and cavity losses. Here we use the same parameters as in Fig.~\ref{fig:exp_comparison}. Colors decode different phases which we characterize by the position of the poles of the cavity photon's Green's function $D_a^R(\omega)$~\eqref{eq:GF_photons}. {Phase (0) is white in the main figure.} Positions of the poles corresponding to different phases are depicted in the right panel (see also Tab.~\ref{tab:phases} for details). Phases (0-1) are stable, phases (2-6) are unstable. The Dicke-type transition obtained from Eq.~\eqref{eq:phase_tr} is shown with a solid line. The ``counter lasing'' instability calculated from Eq.~\eqref{eq:gtic} is shown with an arrow.}
\end{figure*}

The superradiant transition discussed in the previous sections is similar to the phase transition of the Dicke model at thermal equilibrium. For example, the critical exponents of the driven-dissipative model are the same as an equilibrium one at a finite effective temperature~\cite{DallaTorre2013,DallaTorre2016}. An important question is whether this model can display properties that have no equilibrium counterpart~\cite{cheung2017emergent}. In this section, we identify one instance of a genuine non-equilibrium effect, namely a dynamical instability of the system, which cannot be mapped to a Dicke transition.

Dynamical instabilities can be studied by observing the position of the poles of the dressed Green's function of the cavity, $D_a^R(\omega)$. The function is given by Eq.~\eqref{eq:GF_photons} and has four poles. Note that by the construction of the Green's function in Nambu space, the poles either occur on the imaginary axis or come in pairs with the same imaginary part and opposite real parts, $\omega_p \leftrightarrow - \omega_p^*$. { A phase is stable if} the imaginary parts of all the poles are positive, such that $G^{R}_a\left(t-t' \rightarrow \infty \right) \rightarrow 0$. This property leads to two fundamentally different types of instabilities, depending on the number of poles that cross the real axis. In the Dicke transition, a single, pure imaginary pole crosses the origin of the complex plane. Alternatively, one can have a pair of poles that (contemporarily) cross the real axis, giving rise to a distinct type of instability.

\begingroup
\begin{table}[t]
\caption{\label{tab:phases} Phases of the generalized Dicke model classified according to the position of the poles of the retarded Green's function of the cavity photon $D_a^R(\omega)$~\eqref{eq:GF_photons}. For fixed $g$ and $g'$ the Green's function $D_a^R(\omega)$ has four different poles.  }
\centering
\begin{ruledtabular}
\begin{tabular}{ p{0.2cm} p{1.5cm} p{6cm}}
0 & normal & $\text{Re}\left[\omega_\alpha\right]\neq 0$, $\text{Im}\left[\omega_\alpha \right] <0$ for all $\alpha$ \\
\hline
1 & damped & $\text{Re}\left[\omega_{1,2}\right]\neq 0$, $\text{Im}\left[\omega_{1,2} \right] <0$, $\text{Re}\left[\omega_{3,4}\right] =  0$, $\text{Im}\left[\omega_{3,4} \right] <0$ \\
\hline
2 & unstable oscillatory  &  $\text{Re}\left[\omega_{1,2}\right]\neq 0$, $\text{Im}\left[\omega_{1,2} \right] <0$, $\text{Re}\left[\omega_{3,4}\right] \neq 0$, $\text{Im}\left[\omega_{3,4}\right] >0$\\
\hline
3 & unstable & $\text{Re}\left[\omega_{1,2}\right]\neq 0$, $\text{Im}\left[\omega_{1,2} \right] <0$, $\text{Re}\left[\omega_{3,4}\right] = 0$,  $\text{Im}\left[\omega_{3,4}\right] >0$\\
\hline
4 & unstable  &  $\Re\left[\omega_{1,2}\right] = 0$, $\Im\left[\omega_{1,2}\right] <0$, $\Re\left[\omega_{3,4}\right] = 0$, $\Im\left[\omega_{3,4}\right] > 0$  \\ \hline
5 & unstable superradiant & $\Re\left[\omega_{1,2,3}\right] = 0$, $\Im\left[\omega_{1,2,3}\right] <0$, $\Re\left[\omega_{4}\right] = 0$, $\Im\left[\omega_4\right] > 0$\\
\hline
6 & unstable superradiant & $\text{Re}\left[\omega_{1,2}\right]\neq 0$, $\text{Im}\left[\omega_{1,2} \right] <0$, $\text{Re}\left[\omega_{3}\right] =  0$, $\text{Im}\left[\omega_{3} \right] <0$,  $\text{Re}\left[\omega_{4}\right] =  0$, $\text{Im}\left[\omega_{4} \right] >0$\\
\hline
\end{tabular}
\end{ruledtabular}
\end{table}
\endgroup

Fig.~\ref{fig:diag2}(a) shows the complete phase diagram of the generalized Dicke model with dissipation. This phase diagram demonstrates several different phases including the normal and superradiant phases. {See also Table~\ref{tab:phases} for the characterization of all phases. The black solid line represents the points where the the Green's function has a pole at zero frequency, Eq. (37). Note that this line can either separate a stable phase from an unstable one (like in the case of the Dicke transition {between phases 1 and 6}), or two unstable phases (see the upper part of Fig.~\ref{fig:diag2}(a), where the black line separates the unstable phases 5 and 4).}

In addition to the Dicke transition, the present driven-dissipative Dicke model shows {a second instability line between phase 0 (white area, stable) and phase 2 (light-blue area, unstable). As shown in Fig. 6(b), this transition involves the simultaneous transition of two poles across the real axis.}

In order to gain a physical understanding of this instability, we now derive analytic expressions for the instability in two limiting cases: (i) no rotating terms $g = 0$; (ii) zero dissipation $\Gamma = 0$ case. {In both cases, we find simple analytic expressions for the transition, by looking for the point where the imaginary part of the relevant eigenvalues vanishes. { As we will see, the instability occurs when counter-rotating terms overcome the cumulative effect of dissipation and rotating terms, and will be referred to as a ``counter-lasing'' transition. Signatures of this transition were recently observed in the experiments of Ref. \cite{Zhiqiang2017}, and are presented in the next section.}

We start with the description of the counter-lasing instability with finite dissipation and $g=0$. This instability has been observed in~\cite{zhiqiang2018dicke} and referred to as a single beam threshold.
In this case, the transition occurs at the critical coupling
\begin{equation} \label{eq:gtic}
        g'_\text{crit}=
                \sqrt{-\frac{\Gamma\kappa\left(\left(\Gamma+\kappa\right)^2+\left(\omega_0+\omega_c\right)^2\right)}{2s_z\left(\Gamma+\kappa\right)^2}}
\end{equation}
where we are assuming that $s_z<0$. {This point is indicated by an arrow in Fig. \ref{fig:diag2}(a).}

To understand the nature of the transition, let us now focus on the case of a small cavity decay $\Gamma \gg \kappa$, and assume that all the atoms are initially polarized down $s_z=-1/2$. The instability can be easily understood by considering a single atom coupled to the cavity via the interaction $\hat H'=2 g\left(\hat S^+ \hat a+ \hat S^-a^\dagger\right)+2 g'\left(\hat S^+ \hat a^\dagger +\hat S^- \hat a\right)$. Because the model does not have rotating terms, photons can be created only by the term $g' S^+ a^\dagger$. According to Fermi Golden's rule, the rate of this process is $(g')^2\rho_a(\omega_0)$, where $\rho_a(\omega_0) = \Im[1/(\omega_0-\omega_c+i\Gamma)]$ is the atomic density of states. The system becomes unstable when this rate is larger than the photon decay rate $\kappa$, or
\begin{eqnarray}
 g'_\text{crit} = \sqrt{-\frac{\kappa\left(\Gamma^2+\left(\omega_0+\omega_c\right)^2\right)}{2s_z\Gamma}}&,\text{ for }\Gamma\gg\kappa.
\end{eqnarray}
This expression is indeed the limit of Eq.~\eqref{eq:phase_tr} for $\Gamma\gg\kappa$. This instability is equivalent to a lasing transition, where the rate of photon generation becomes larger than the rate of photon decay. Unlike the usual lasing transition, the present instability is driven by counter-rotating terms.}

Let us now consider the case of zero atomic dissipation $\Gamma=0$, where the instability occurs at
\begin{equation}\label{eq:udic}
\frac{g}{g'}=\sqrt{1-\frac{4\omega_0\omega_c}{\kappa^2+\left(\omega_0+\omega_c\right)^2}}.
\end{equation}

If we assume the cavity to be initially empty (Markovian bath), only two terms of $\hat H'$ can act on the state, namely $2 g\hat S^- \hat a^\dagger$ and $2 g' \hat S^+\hat a^\dagger$.
These two terms respectively flip the spin from down to up and vice versa.
Their rates are respectively given by $\gamma_{\text{eff},\downarrow} = (g')^2 \rho_a(\omega_c)$ and $\gamma_{\text{eff},\uparrow} = g^2 \rho_a(-\omega_c)$, where $\rho_a=\Im[1/(\omega-\omega_0+i \kappa)]$ is the density of states of the cavity.
The system becomes unstable when the effective flip rate upwards is larger than the downwards flip rate.
Thus, this instability occurs when $\gamma_{\text{eff},\uparrow}=\gamma_{\text{eff},\downarrow}$. {This condition is equivalent to Eq.~\eqref{eq:udic}.}

\subsection{Comparison with experiments}
\label{sec:experiment}

\begin{figure}[t]
\begin{centering}
\includegraphics[width=0.9\columnwidth]{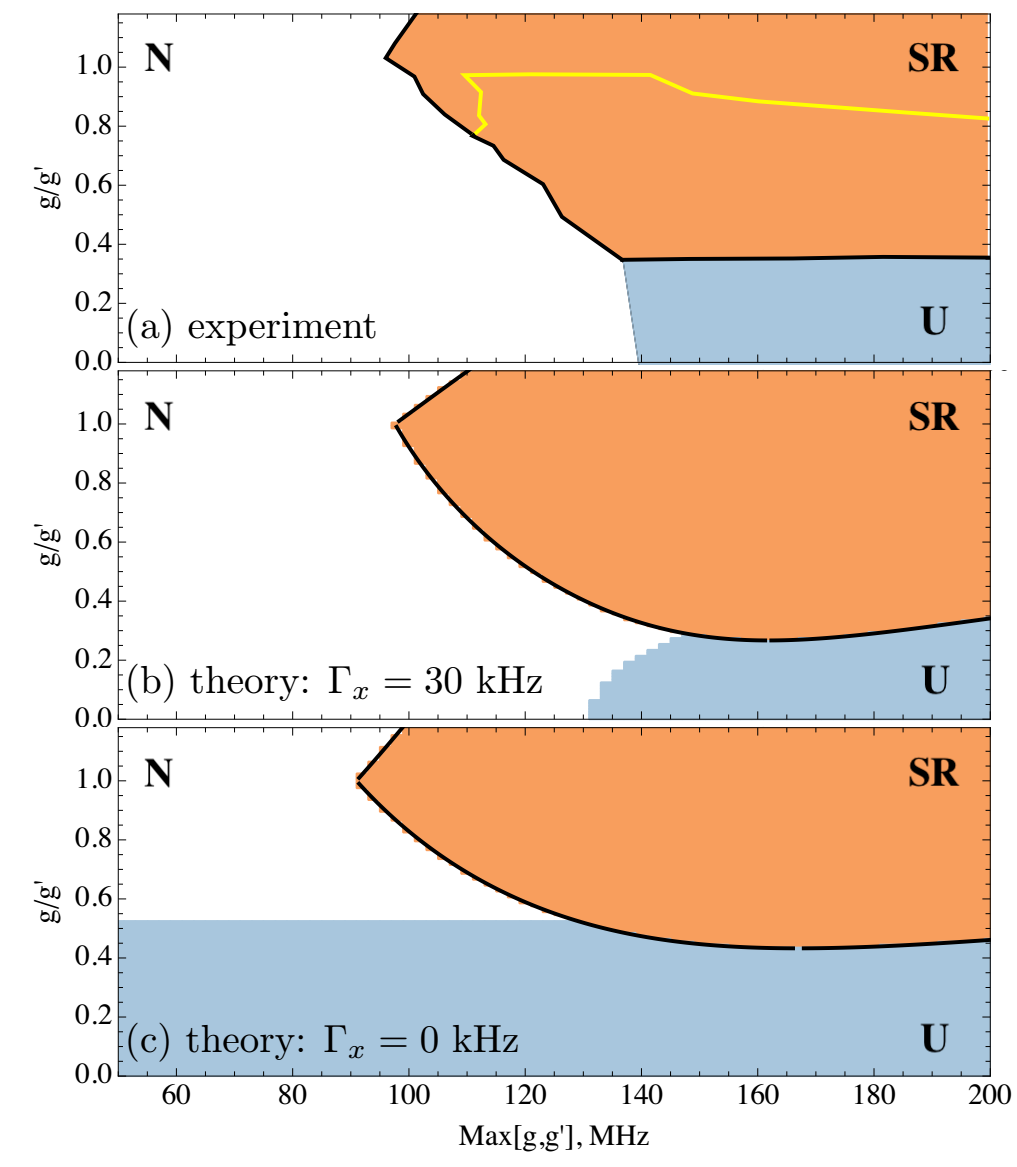}
\par\end{centering}
\caption{\label{fig:exp_comparison} Comparison between (a) experimental and (b-c) theoretical phase transition diagrams for the generalized Dicke system. N, SR, and U denotes normal, superradiant, and unstable phases correspondingly. Parameters used for theoretical calculation correspond to the experimental data: cavity mode frequency $\omega_c=100$~kHz, dissipation $\kappa=107$~kHz, and energy splitting $\omega_z = 77.2$ kHz. In theoretical calculations, we consider that the atomic interaction with the incoherent bath partially polarizes the system, $s_z = -0.25$, and the strength of dissipation is $\Gamma =30$~kHz (b) and $\Gamma=0$~kHz (c). }
\end{figure}

We now compare the result of our calculation with a recent experimental realization of the generalized Dicke model using a gas of ultracold  $^{87}$Rb atoms confined to a high finesse cavity~\cite{Zhiqiang2017}. The atomic system used in the experiment has a multi-level structure which is similar to the four-level scheme considered in this paper (see the sketch in Fig.~\ref{fig:Sketch_Spin}(b)). The main difference is that, after adiabatic elimination, the experimental system maps onto the spin-1 generalized Dicke model, while in this paper we consider a system where the atoms are effectively described as spin-$\frac{1}{2}$ models. As the atoms are highly polarized in the normal phase, we expect the spin-$1$ and spin-$\frac{1}{2}$ models to behave similarly.

Fig.~\ref{fig:exp_comparison} shows the comparison between the experimentally observed phase diagram and the theoretical calculations {with and without atomic dissipation}. {The choice of the somewhat un-natural axes (``$g/g'$'' {\it vs.} ``$\text{max}(g,g')$'') is determined by the details of the experimental protocol, in which $g$ and $g'$ are adiabatically turned on at a fixed ratio (i.e. along the horizontal lines of Fig. ~\ref{fig:exp_comparison}). The threshold to instability was experimentally determined as the value of the parameters at which a jump in the number of photons was observed.} { The experimentally observed phase diagram (Fig.~\ref{fig:exp_comparison}(a)) includes three distinct regions, which we identify with the normal  (white), the super-radiant (orange), and the so-called ``counter-lasing'' (light-blue) phases~\cite{Zhiqiang2017}.}

Our calculations demonstrate the importance of the single atom decay and dephasing channels, modeled by $\Gamma$.
In Fig.~\ref{fig:exp_comparison}(b) and (c), we compare the theoretical predictions without dissipation, $\Gamma= 0$~kHz, and with a weak atomic dissipation, $\Gamma= 30$~kHz. According to Eq.~\eqref{eq:udic}, in the absence of dissipation, the region where $g/g' < 0.53$ is unstable, for any value of $g'$. A similar result was obtained in~\cite{Zhiqiang2017}. {This theoretical prediction is inconsistent with the experimental findings, which found a sudden jump in the number of photons at a finite value of $g'$.} Fig.~\ref{fig:exp_comparison}(b) shows that when dissipation is present in the system, the ``counter-lasing'' transition occurs at a finite value of $g'$. Moreover, the instability threshold shown in Fig.~\ref{fig:exp_comparison}(b) allows us to indirectly access the microscopic parameters of the model using Eq.~\eqref{eq:gtic}. Our theory provides the best fit to the experimental results with $\Gamma= 30$~kHz.

\section{Comparison with Lindblad Master equations}
\label{sec:Lindblad}
\subsection {Spin-spin correlations of a dissipative two-level systems}
We now use the Lindblad master equation to compute the spin-spin
correlation functions for a two-level system with decay rates $\gamma_{\downarrow}$ and $\gamma_\uparrow$. As we will show, the results of this approach are the same as those obtained in Sec.~\ref{sub:spin_prop} using fermionic path integrals.
According to the quantum regression theorem \cite{breuer2002theory}, in the case of Markovian master equations, the spin-spin correlation functions can be directly computed from the evolution of the spin operators. In the absence of spin-cavity coupling ($g=g'=0$), one obtains
\begin{eqnarray}
\label{eq:spin_op}
{\hat S_i^z(t)}&=&e^{-2\Gamma\left(t-t'\right)}\left(  {\hat S_i^z(t')}-s_z\right)+s_z,\\ \nonumber
{\hat S_i^x(t)}&=&e^{-\Gamma\left(t-t'\right)} \biggl(  {\hat S_i^x(t')} \cos{\bigl(\omega_0 \left(t-t'\right)\bigr)}\biggr.\\ \nonumber
&&\biggl.-\hat S_i^y(t') \sin{\bigl(\omega_0 \left(t-t'\right)\bigr)}\biggr),
\end{eqnarray}
To obtain the spin-spin correlation functions it is now sufficient to multiply both sides of Eqs.~\eqref{eq:spin_op} by the relevant spin operator at time $t'$.
Finally, by using the identities $\av{\hat S_i^n(t')\hat S_i^n(t')}=\frac14$, $\av{\hat S_i^n(t')\hat S_i^m(t')}=\frac12i\ve_{nml}\av{\hat S_i^l(t')}$,  and $\av{\hat S_i^z(t')}=s_z$, we find:
\begin{eqnarray}
\label{eq:lin-corr}
\av{\hat S_i^z(t)\hat S_i^z(t')}&=&
\frac{e^{-2\Gamma \left(t-t'\right)}}4\\ \nonumber
&&+s_z^2\left(e^{-2\Gamma\left(t-t'\right)}-1\right),\\ \nonumber
\av{\hat S_i^x(t)\hat S_i^x(t')}&=&
\frac{e^{-\Gamma\left(t-t'\right)}}4\biggl(\cos\bigl(\omega_0\left(t-t'\right)\bigr) \biggr.\\\nonumber
&&\biggl.-2is_z\sin\bigl(\omega_0\left(t-t'\right)\bigr)\biggr).
\end{eqnarray}
This expression agrees with the diagrammatic approach, Eqs.~(\ref{eq:sxsx_final}) and (\ref{eq:szsz_final}) (shown in Fourier space).

\subsection{Lindblad master equation of the generalized Dicke model}
In this section, we compare our results obtained using Keldysh formalism with the predictions of
the Lindblad approach.
Our starting point is the generalized Dicke model, Eq.~\eqref{eq:gDM}.
We focus on the steady-state of the system, induced by the interplay between this Hamiltonian and the dissipative channels associated with the cavity decay $\kappa$, and the single-atom losses $\gamma_{\uparrow/\downarrow}$.
In the rotating frame, the system can be described by the Lindblad master equation:
\begin{eqnarray}\label{eq:lin-lin}
\frac{\mathrm{d} \hat \rho}{\mathrm{d} t} &=&
        -i\left[\hat  H, \hat\rho\right]
        +\kappa\mathcal{D}\left[ \hat a\right] \\ \nonumber
        &+&\sum_i^N\left(
        \gamma_\uparrow\mathcal{D}\left[\hat S_i^+\right]
        +\gamma_\downarrow\mathcal{D}\left[\hat S_i^-\right]
        \right),
\end{eqnarray}
where $\mathcal{D}\left[\hat x\right] \equiv 2 \hat x \hat \rho \hat x^\dagger-\left\{\hat x^\dagger \hat x, \hat \rho\right\}$.
Following Ref.~\cite{Kirton2017}, we first use Eq.~\eqref{eq:lin-lin} to derive the equations for the
collective variables
$\hat a$, $\hat S^x$, $\hat S^y$ and $\hat S^z$, where $\hat S^n\equiv\sum_i^N S_i^n/N$.
Using the commutation relations $\left[\hat a,\hat a^\dagger\right]=1$ and $\left[\hat S^n,\hat S^m\right]=i\ve_{nml}\hat S^l/N$, we obtain:

\begin{eqnarray} \label{eq:lin-a}
\partial_t{\hat a}&=& -\left(i\omega_c+\kappa\right) {\hat a}\\ \nonumber &&
                -{i\left(g+g'\right)}\sqrt{N}\hat S^x
                +{\left(g'-g\right)}{\sqrt{N}} {\hat S^y}, \\  \nonumber
\partial_t  {\hat S^x}&=&
        -\omega_0  {\hat S^y}-\sfrac{i\left(g'-g\right)}{\sqrt{N}}  {\hat S^z}\left(  {\hat a}- {\hat a^\dagger}\right)
                -\Gamma {\hat S^x}, \\  \nonumber
\partial_t  {\hat S^y}&=&
        \omega_0  {\hat S^x}-\sfrac{\left(g+g'\right)}{\sqrt{N}}  {\hat S^z} \left(  {\hat a}+  {\hat a^\dagger}\right)
                -\Gamma {\hat S^y},\\ \nonumber
\partial_t  {\hat S^z}&=&
                -2\gamma_\downarrow\left(\frac12+  {\hat S^z}\right)
                +2\gamma_\uparrow\left(\frac12-  {\hat S^z}\right)\\ \nonumber &&
                +\sfrac{\left(g+g'\right)}{\sqrt{N}}  {\hat S^y}\left(  {\hat a}+  {\hat a^\dagger}\right)
                +\sfrac{i\left(g'-g\right)}{\sqrt{N}}  {\hat S^x}\left(  {\hat a}-  {\hat a^\dagger}\right),
\end{eqnarray}
where $\Gamma = \gamma_\uparrow+\gamma_\downarrow$.

\par We now determine the phase diagram of the model by studying the linear
stability of Eqs.~\eqref{eq:lin-a},
around their normal phase, defined by
$\av{\hat a}=\av{\hat S^x}=\av{\hat S^y}=0$. { The resulting equations of motions are best described in terms of the vector $\delta \hat R^T =\left( \delta \hat a, \delta \hat a^\dagger,
\delta \hat  S^x, \delta \hat  S^y, \delta \hat  S^z \right) $,
where we have defined
$\delta\hat S^\alpha=\hat S^\alpha-\av{\hat S^\alpha}$,
$\hat\delta a=\hat a-\av{\hat a}$.
 Up to first order in $\delta \hat R$, Eqs.~(\ref{eq:lin-a}) lead to}
\begin{eqnarray}
        \label{eq:lleom}
        -i\partial_t \delta\hat R& =& M \delta \hat R,
\end{eqnarray}
where the linear response matrix $M$ is defined by

\begin{align} \label{eq:lin-m-def}
        M=&
        \left(\begin{smallmatrix}
                -\omega_c+i \kappa & 0 & - {\left(g+g'\right)}{\sqrt{N}} &{i \left(g-g'\right)}{\sqrt{N}} & 0 \\
                0 & \omega_c+i\kappa &     {\left(g+g'\right)}{\sqrt{N}}& {i\left(g-g'\right)}{\sqrt{N}} & 0\\
                 \frac{\left(g-g'\right)}{\sqrt{N}}s_z & -\frac{\left(g-g'\right)}{\sqrt{N}}s_z & i\Gamma& i\omega_0 & 0\\
                \frac{i\left(g+g'\right)}{\sqrt{N}}s_z & \frac{i\left(g+g'\right)}{\sqrt{N}}s_z &-i\omega_0 & i\Gamma & 0\\
                0&0&0&0&2i\Gamma
        \end{smallmatrix}\right),
\end{align}
{and $s_z \equiv \av{S^z_i} = \av{\hat S^z}$ is given by Eq.~(\ref{eq:spin_exp}).}

\subsection{The superradiant transition}
Following the analysis in Sec.~\ref{sub:phase_tr} we define the superradiant transition with the condition that one of the eigenvalues of $M$ is exactly equal to zero. Equivalently this condition can be written as $\det\left(M\right)=0$. Taking the determinant of the matrix $M$, we obtain the condition for the superradiant transition, which is identical to the one obtained using Green's functions, Eq.~\eqref{eq:phase_tr}.

\section{Conclusions and Outlook}
\label{sec:concl}

In this paper, we presented a fermionic path-integral analysis of driven-dissipative atomic systems.
Our goal was to demonstrate that the Keldysh diagrammatic technique is suitable for the analysis of many-body atomic systems that have a multi-level structure and { interact with both coherent and incoherent photonic modes}.

First, we showed that the Majorana fermion representation can simplify the calculation of spin-spin correlation functions. We demonstrated that, when the expectation value of the spin in the steady state is zero ($\av{S^\alpha} =0$), the corresponding spin-spin correlation function ($\av{S^\alpha(t) S^\alpha(t')}$) can be calculated as a single Green's function. This is in contrast with the case where the expectation value is finite ($\av{S^\alpha} \neq 0 $) and the corresponding spin-spin correlation function needs to be calculated from the convolution of two Green's functions. {This observation explains contradictory results reported in the literature.}

Next, we demonstrated that the fermionic language is natural for the description of the interaction of an atomic system with an incoherent dissipative bath when the system is driven by an external field. We showed that the adiabatic elimination of the far detuned states can be done with the help of Gaussian integrals. {We specifically considered situations were the far-detuned states do not introduce any retardation effects.} In the general case, those effects can be naturally included with the use of non-equilibrium field theory and can lead to additional non-Markovian correlations in the effective bath.

After considering the impact of dissipation on a single atom, we extended our formalism to the case of a $N$-atom system inside of an optical resonator. We considered the case where the atoms are pumped with an external field and interact with a dissipative environment. By analogy with the results for a single driven atom interacting with a dissipative bath, we introduced an effective Hamiltonian for this problem and used a diagrammatic technique to describe the steady state of the system. We classified the self-energy contributions according to their scaling with the effective coupling strength $g=\lambda\sqrt{N}$. In particular, we observed that the back action of the cavity on the spin system scales as $\lambda^2=g^2/N$ and can be neglected in the leading order approximation.

We described the instabilities of the system using the Green's function language. The Dicke transition is signaled by a pole of the Green's function approaching zero frequency. {In contrast, when two conjugated poles simultaneously cross the real axis, the system shows a distinct type of instability}. One example is given by the ``counter-lasing'' instablity, which occurs when the counter-rotating terms overcome the atomic and photonic decay channels. We highlighted the nature of this instability by considering some limiting cases, where its position could be determined based on simple physical arguments.

We compared our theoretical prediction with the experimental observation of the Dicke phase transition  by Ref.~\cite{Zhiqiang2017} in a cavity QED system. Our analysis offers a better description of the experimental situation when compared to the previous analysis, which neglected single-atom decay channels. We demonstrated the importance of atomic dissipation by comparing the experiment to the theoretical calculations with and without dissipation. Moreover, we conclude that the ``counter lasing'' instability was observed experimentally and the instability threshold allows us to indirectly access the microscopic parameters of the model. Our theory {suggests that the experimental results of~\cite{Zhiqiang2017} are best described using $\Gamma= 30$~kHz.}

Lastly, we showed that our theoretical results are in agreement with Lindblad master equation calculations. We showed how our results obtained using diagrammatic formalism translates to the language of master equations. Our study demonstrates the applicability of fermionic path integrals to multi-level atomic systems. This result opens the way to the discussion of non-Markovian dissipative baths, higher order corrections from atom-cavity interactions, interactions between atoms, and effects of disorder. Importantly, the present path-integral approach is not limited to steady state configurations. To study the real-time dynamics of the model it is sufficient to consider Green's functions that depend on two times. Their time evolution is determined by the Kadanoff-Baym equations (see Appendix~\ref{sec:app3}), which need to be solved self-consistently. This approach allows to take into account time evolution and retardation on equal footing.

\begin{acknowledgments}We acknowledge useful discussions with Pjotrs Grisins, Valentin Kasper, Jamir Marino, Giovanna Morigi, Florentin Reiter, Kushal Seetharam, and Eli Wilner. This work is supported by Harvard-MIT CUA, NSF Grant No. DMR-1308435, AFOSR Quantum Simulation MURI, AFOSR grant number FA9550-16-1-0323. E.G.D.T and M.M.R. are supported by the Israeli Science Foundation Grant No. 1542/14.\end{acknowledgments}

{\it Note: after the submission of this manuscript, we became aware of an independent study \cite{kirton2017supperradiant}, where the counter-lasing transition was discussed.}

{Y.S and M.M.R equally contributed to this work.}

\begin{appendix}

\section{Derivation of the spin dissipation}
\label{sec:app1}
In this Appendix, we provide a detailed calculation of the self-energies of the $\hat f$- and $\hat \eta$- fermions from Sec.~\ref{sec:diagr}.
The diagrams shown in Fig.~\ref{fig:SE-2-3} correspond to the interaction of fermions with an effective dissipative bosonic environment.

In order to evaluate the diagrams, we recall the definitions of the spectral functions of the bath in the rotating frame, which we previously introduced in the main text,
\begin{eqnarray} \label{eq:bath_properties_app}
&& \sum_{k}\frac{ i\Omega_L^2 \lambda_{k,L}^{2} }{ 8 \Delta_L^2 }  D_{k,L}^{>}(\omega)	 =   \gamma_\uparrow, \\ \nonumber
&& \sum_{k}\frac{i \Omega_L^2 \lambda_{k,L}^{2} }{ 8 \Delta_L^2 }  D_{k,L}^{<}(\omega) 	= 0, \\ \nonumber
&& \sum_{k}\frac{ i\Omega_R^2 \lambda_{k,R}^{2} }{ 8 \Delta_R^2 } D_{k,R}^{>}(\omega) =   \gamma_\downarrow,\\ \nonumber
&& \sum_{k}\frac{ i\Omega_R^2 \lambda_{k,R}^{2} }{ 8 \Delta_R^2 } D_{k,R}^{<}(\omega) = 0.
\end{eqnarray}
Note that the sum of the lesser Green's functions of the bosons is zero. This is the result of the Markovian approximation.

\paragraph{$\hat f$-fermion self-energy. -- }
First, we calculate the self-energy of the $\hat f$-fermion. Using the Langreth rules, we write the greater and lesser parts of the self-energy function:
\begin{eqnarray}
\label{eq:sigma_f_Gr_app} \nonumber
\Sigma_f^> (\omega) &&=\frac{i}{2}  \sum_k \frac{\Omega_R^2}{\Delta_R^2} \lambda_{k,R}^2  \int \frac{d\varepsilon}{2\pi}
G_\eta^>( \varepsilon)  D^<_{k,R} (\omega - \varepsilon) \\
+&&\frac{i}{2}  \sum_k \frac{\Omega_L^2}{\Delta_L^2} \lambda_{k,L}^2  \int \frac{d\varepsilon}{2\pi}
G_\eta^> (\varepsilon) D^>_{k,L} (\omega - \varepsilon)   \\
\label{eq:sigma_f_Ls_app} \nonumber
\Sigma_f^< (\omega) &&= \frac{i}{2}  \sum_k \frac{\Omega_R^2}{\Delta_R^2} \lambda_{k,R}^2  \int \frac{d\varepsilon}{2\pi}
G_\eta^< (\varepsilon)  D^>_{k,R} (\omega - \varepsilon)  \\
+&&\frac{i}{2}  \sum_k \frac{\Omega_L^2}{\Delta_L^2} \lambda_{k,L}^2  \int \frac{d\varepsilon}{2\pi}
G_\eta^< (\varepsilon)  D^<_{k,L} (\omega - \varepsilon).
\end{eqnarray}

We calculate the retarded and Keldysh components of the self-energy by adding and subtracting greater and lesser self-energies
\begin{eqnarray}
\Sigma_f^R (\omega) - \Sigma_f^A(\omega) &=& \Sigma_f^>(\omega) - \Sigma_f^<(\omega),\\ \nonumber
\Sigma_f^K (\omega) &=& \Sigma_f^>(\omega) + \Sigma_f^<(\omega).
\end{eqnarray}

Substituting Eq.~\eqref{eq:sigma_f_Gr_app} we obtain
\begin{eqnarray} \nonumber
&&\Sigma_f^R (\omega) -\Sigma_f^A(\omega) =
2\gamma_{\downarrow}   \int \frac{d\varepsilon}{\pi}
G_\eta^> (\varepsilon)  -
2 \gamma_{\uparrow} \int \frac{d\varepsilon}{\pi}  G_\eta^< (\varepsilon), \\
&& \Sigma_f^K (\omega)  =
2\gamma_{\downarrow}   \int \frac{d\varepsilon}{2\pi}
G_\eta^> (\varepsilon)  +
 2 \gamma_{\uparrow} \int \frac{d\varepsilon}{\pi}  G_\eta^< (\varepsilon).
\end{eqnarray}

By definition, the greater and lesser Majorana Green's functions are not independent, $G_\eta^>(\omega) = -G_\eta^<(-\omega)$. The integral over the greater Majorana functions is a constant $2 \int \frac{d\varepsilon}{\pi} G_\eta^> (\varepsilon) = - i$. Thus the self-energies read
\begin{eqnarray} \label{eq:selfenergy_A}\nonumber
\Sigma_f^R (\omega)   - \Sigma_f^A(\omega)
 &=& - i (\gamma_{\downarrow}  + \gamma_{\uparrow}),  \\
 \Sigma_f^K (\omega)
 &=& - i(\gamma_{\downarrow}  - \gamma_{\uparrow}).
\end{eqnarray}

Using the self-energies~\eqref{eq:selfenergy_A}, we calculate the ratio between them (assuming $\Sigma_f^K (\omega) \neq 0$ and $\Sigma_f^R (\omega)- \Sigma_f^A (\omega) \neq 0$)
\begin{equation} \label{eq:occup_f}
h_f (\omega) = \frac{\Sigma_f^> (\omega)+\Sigma_f^> (\omega)}{\Sigma_f^> (\omega)- \Sigma_f^<(\omega) } =\frac{\gamma_\downarrow - \gamma_\uparrow}{\gamma_\downarrow + \gamma_\uparrow}.
\end{equation}
In the stationary state, the solution of the Dyson equation for greater and lesser Green's functions reads
\begin{eqnarray}
G_f^{>}  (\omega)  &=& G_f^R  (\omega)  \Sigma_f^{>}  (\omega)  G_f^A  (\omega),  \\ \nonumber
G_f^{<}  (\omega)  &=& G_f^R  (\omega)  \Sigma_f^{<}  (\omega)  G_f^A  (\omega).
\end{eqnarray}
We multiply numerator and denominator by $G_f^R$ (from the left) and $G_f^A$ (from the right), and obtain the expression that connects the greater and lesser Green's functions in the stationary state (also known as the non-equilibrium fluctuation-dissipation relation):
\begin{equation}
h_f (\omega) = \frac{G_f^>(\omega) + G_f^<(\omega) }{G_f^>(\omega) - G_f^<(\omega)  }.
\end{equation}

Summing up, the expressions for the greater and lesser $\hat f$-fermion Green's functions are
\begin{eqnarray}\label{eq:Gf_f_gr_and_ls}
 G_f^>(\omega) &=& -i\pi (1 + h_f (\omega) )\rho_f(\omega),\\ \nonumber
 G_f^<(\omega) &=& i\pi (1 -  h_f (\omega) )\rho_f(\omega), \\ \nonumber
 \rho_f(\omega) &=& \frac{1}{\pi} \frac{\gammau + \gammad}{\left( \omega-\omega_z\right)^2 + \left( \gammau+\gammad\right)^2}.
\end{eqnarray}

\paragraph{$\hat \eta$-fermion self-energy.--}
We now examine the self-energy of the $\hat \eta$-fermion. Using the Langreth rules we write the greater and lesser components of the self-energies
\begin{eqnarray}
\label{eq:sigma_eta_Gr_app} \nonumber
\Sigma_\eta^> (\omega) &&= i  \sum_k \frac{\Omega_R^2}{\Delta_R^2} \lambda_{k,R}^2  \int \frac{d\varepsilon}{2\pi}
G_f^>( \varepsilon)  D^>_{k,R} (\omega - \varepsilon) \\
+&& i  \sum_k \frac{\Omega_L^2}{\Delta_L^2} \lambda_{k,L}^2  \int \frac{d\varepsilon}{2\pi}
G_f^> (\varepsilon) D^<_{k,L} (\omega - \varepsilon)   \\
\label{eq:sigma_eta_Ls_app} \nonumber
\Sigma_\eta^< (\omega) &&= i  \sum_k \frac{\Omega_R^2}{\Delta_R^2} \lambda_{k,R}^2  \int \frac{d\varepsilon}{2\pi}
G_f^< (\varepsilon)  D^<_{k,R} (\omega - \varepsilon)  \\
+&& i \sum_k \frac{\Omega_L^2}{\Delta_L^2} \lambda_{k,L}^2  \int \frac{d\varepsilon}{2\pi}
G_f^< (\varepsilon)  D^>_{k,L} (\omega - \varepsilon)
\end{eqnarray}

We calculate the retarded and Keldysh components of the self-energy:
\begin{eqnarray}
\Sigma_\eta^R (\omega)-\Sigma_\eta^A (\omega)  &=& \Sigma_\eta^>(\omega) - \Sigma_\eta^<(\omega)\\ \nonumber
\Sigma_\eta^K (\omega) &=& \Sigma_\eta^>(\omega) + \Sigma_\eta^<(\omega)
\end{eqnarray}

Substituting~\eqref{eq:sigma_f_Gr_app} we obtain
\begin{eqnarray}\nonumber
\Sigma_\eta^R (\omega) -\Sigma_\eta^A (\omega) &=&
4 \gamma_{\uparrow}   \int \frac{d\varepsilon}{\pi}
G_f^> (\varepsilon)  -
4 \gamma_{\downarrow} \int \frac{d\varepsilon}{\pi}  G_f^< (\varepsilon) \\ \nonumber
 \Sigma_\eta^K (\omega)  &=&
4\gamma_{\uparrow}   \int \frac{d\varepsilon}{\pi}
G_f^> (\varepsilon)  +
4 \gamma_{\downarrow} \int \frac{d\varepsilon}{\pi}  G_f^< (\varepsilon)
\end{eqnarray}

Substituting the Green's functions of the $\hat f$-fermion~\eqref{eq:Gf_f_gr_and_ls}, we obtain
\begin{eqnarray}
\Sigma_\eta^R (\omega) -\Sigma_\eta^A (\omega)  &=&
-2 i \left( (\gamma_{\uparrow} -  \gamma_{\downarrow} ) h_f  +
(\gamma_{\uparrow} +  \gamma_{\downarrow} ) \right)  \\ \nonumber
&=&-2 i (\gamma_{\uparrow} +  \gamma_{\downarrow} ) \left(1- \left( \frac{\gamma_\downarrow - \gamma_\uparrow}{\gamma_\downarrow + \gamma_\uparrow}\right)^2
\right)  \\ \nonumber
 \Sigma_\eta^K (\omega)  &=&
- 2 i \left( (\gamma_{\uparrow} +  \gamma_{\downarrow} ) h_f +
(\gamma_{\uparrow} -  \gamma_{\downarrow} )   \right)  = 0
\end{eqnarray}

\section{Spin-spin correlation functions}
\label{sec:app2}
We now provide a detailed calculation of the spin-spin correlation functions $\av{S^x (t) S^x (t')}$ and $\av{S^z (t) S^z (t')}$ introduced in Sec.~\ref{sub:spin_prop}.
First, we express the effective spin correlation functions in terms of fermions using Eq.~\eqref{eq:spin_spin_fermions} in the main text. And then, using the equations for the greater and lesser Green's functions Eq.~\eqref{eq:Gf_f_gr_and_ls}, we simplify those expressions.

\paragraph{$\av{S^x (t) S^x (t')}$ correlations. --}
We express the correlation function using $\hat f$ and $\hat \tau_x$ fermions. As we discussed in the main text, it simplifies the calculation.
\begin{eqnarray} \label{eq:sxsx_B1}
\av{S^x (t) S^x (t')} &=& \frac{1}{4}\left\langle\tau_x(t) \left( \hat f (t) +\hat f^\dag (t) \right) \times  \right. \\ \nonumber
&& \left.\tau_x(t')  \left( \hat f (t') + \hat f^\dag (t') \right)\right\rangle
\end{eqnarray}
We use Wick's theorem and express Eq.~\eqref{eq:sxsx_B1} in terms of Green's functions.
\begin{eqnarray} \label{eq:sxsx_B2}
\av{S^x (t)\sigma^x (t')} &=& \frac{i}{4} G_f^>(t,t') -  \frac{i}{4} G_f^<(t',t)
\end{eqnarray}
We transform Eq.~\eqref{eq:sxsx_B2} to the frequency domain and use the expressions for the $\hat f$-fermion Green's functions Eq.~\eqref{eq:Gf_f_gr_and_ls}.
\begin{multline} \label{eq:sxsx_B3}
\fourier{\av{S^x (t) S^x (t')}} = \frac{i}{4} G_f^>(\omega) - \frac{i}{4} G_f^<(-\omega)  = \\
\frac{\pi }{4} ( 1 + h_f (\omega)) \rho_f(\omega) + \frac{\pi }{4} ( 1 - h_f (-\omega)) \rho_f(-\omega)
\end{multline}
Substituting the expression for $\rho(\omega)$ and $h_f (\omega)$ we obtain
\begin{multline} \label{eq:sxsx_B4}
\fourier{\av{S^x (t) S^x (t')}} =
 \frac{\gammad}{2}\frac{1}{\left( \omega-\omega_z\right)^2 + \left( \gammap\right)^2} +\\
\frac{\gammau}{2}\frac{1}{\left( \omega+\omega_z\right)^2 + \left( \gammap\right)^2}
\end{multline}
We take the inverse Fourier transform (by closing the integration contour in the lower half plane) and we the expression in the time domain
\begin{multline}
\av{S^x (t+\tau) S^x (t)} = \frac{1}{4} e^{-\gamma _+ \tau }\left(\cos \left(\tau  \omega _z\right) \right. \\ \left.-i\frac{\gamma _{\downarrow }-\gamma _{\uparrow }}{\gamma _{\downarrow }+\gamma _{\uparrow }}\sin \left(\tau  \omega _z\right)\right)
\end{multline}
This expression is equivalent to the Eq.~\eqref{eq:lin-corr} obtained with the Lindblad approach in Sec.~\ref{sec:Lindblad}.

\paragraph{$\av{S^z (t) S^z (t')}$ correlations. --}
We define the correlation function using the $f$-fermions:
\begin{eqnarray} \nonumber \label{eq:szsz_B1}
   \av{S^z (t) S^z (t')} &=&  \av{\left(  f^\dag(t) f(t) - \frac{1}{2} \right) \left(f^\dag(t') f(t') - \frac{1}{2} \right)} \\ \nonumber
     &=& \av{( f^\dag(t) f(t) f^\dag(t') f(t') } -  \frac{1}{2}\av{ f^\dag(t) f(t) } \\
     &-&\frac{1}{2} \av{ f^\dag(t') f(t') }+\frac{1}{4}
\end{eqnarray}
We use Wick's theorem and express Eq.~\eqref{eq:szsz_B1} in terms of Green's functions.
\begin{eqnarray}
\av{S^z (t) S^z (t')} &=& \av{S^z(t)}\av{S^z(t')} \\ \nonumber
&+& G_f^<(t',t) G_f^>(t,t')
\end{eqnarray}
In the steady state, the correlation function depends only on the time difference. Hence, we write it in the frequency domain
\begin{eqnarray}
\fourier{\av{S^z (t) S^z (t')}} &=& 2\pi \av{S^z}^2\delta(\omega) \\ \nonumber
  &+&\int \frac{d\varepsilon}{2\pi}  G_f^<(\varepsilon) G_f^>(\omega+\varepsilon)
\end{eqnarray}
Substituting the expression for the greater and lesser Green's functions~\eqref{eq:Gf_f_gr_and_ls}, we obtain
\begin{eqnarray}
\fourier{   \av{S^z (t) S^z (t')}} &=& 2\pi  \av{S^z}^2\delta(\omega)  \\ \nonumber
   &+& (\pi)^2 (1 - h_f^2) \int \frac{d\varepsilon}{2\pi} \rho_f (\varepsilon) \rho_f (\omega+\varepsilon)
\end{eqnarray}
Taking the integral in the previous equation, we obtain
\begin{multline}
\fourier{   \av{S^z (t)S^z (t')}} = 2\pi \av{S^z}^2\delta(\omega) \\
   + \left(\frac{1}{4} - \av{S^z}^2\right) \frac{4 (\gamma_\uparrow + \gamma_\downarrow) }{\omega^2+4 (\gamma_\uparrow + \gamma_\downarrow)^2}
\end{multline}
We take the inverse Fourier transform (by closing the integration contour in the lower half plane) and obtain the expression in the time domain
\begin{multline}
\left\langle  S^z\left(t+\tau )\right) S^z \left(t\right)  \right\rangle =\left\langle S^z\right\rangle ^2 \\ -e^{-2\left( \gammau + \gammad\right) \tau } \left(\frac{1}{4}-\left\langle S^z\right\rangle^2 \right)
\end{multline}
This expression is equivalent to the Eq.~\eqref{eq:lin-corr} obtained with the Lindblad approach in Sec.~\ref{sec:Lindblad}.

\section{Rotating Wave Approximation}
\label{sec:app4}
In this Appendix we give details on the rotating wave approximation (RWA) presented in section \ref{sec:RWA}.
This transformation will take us from Eqs.~\eqref{eq:H1} and \eqref{eq:orig_haml} to Eq.~\eqref{eq:rwa_haml}.
The RWA consists of two steps: we first apply a unitary transformation to the Hamiltonian (Eqs.~\eqref{eq:H1} and~\eqref{eq:orig_haml}) and then neglect the fast oscillating terms.

Let us begin by examining the Schrodinger equation:
\begin{align}
i\partial_t\ket\psi=&H(t)\ket\psi.
\end{align}
If we will apply a unitary transformation $U(t)$ to the state $\ket\psi$, we obtain a state $\ket{\psi'}=U(t)\ket\psi$, which is governed by the Schroedinger equation.
\begin{align}
\label{eq:new_h}
\begin{split}
i\partial_t\ket{\psi'}=&i\partial_t\left(U(t)\ket\psi\right)\\
=&iU(t)\partial_t\ket\psi+i\partial_tU(t)\ket\psi\\
=&U(t)H(t)\ket\psi+i\dot{U}(t)\ket\psi\\
=&U(t)H(t)U^\dagger(t)\ket{\psi'}+i\dot{U}(t)U^\dagger(t)\ket{\psi'}\\
=&\tilde{H}(t)\ket{\psi'}.
\end{split}
\end{align}
Here we defined a new Hamiltonian $\tilde{H}(t)=U(t)H(t)U^\yd(t)+i\dot{U}(t)U^\yd(t)$.
In section \ref{sec:RWA} we used the following transformation:
\begin{align}
U(t)=&\prod\limits_{i=1}^4U_{c_i}(t)\prod\limits_{k,\sigma}U_{d_{k,\sigma}}(t),
\end{align}
where:
\begin{align}
U_{c_1}(t)=&\mathbb{I\!\!I},\\
\nonumber
U_{c_2}(t)=&e^{i\frac{\omega_L-\omega_R}2c_2^\yd c_2^\nd t},\\
\nonumber
U_{c_3}(t)=&e^{i\omega_{dr}c_3^\yd c_3^\nd t},\\
\nonumber
U_{c_4}(t)=&e^{i\omega_Lc_4^\yd c_4^\nd t},\\
\nonumber
U_{d_{k,\sigma}}(t)=&e^{i\omega_{dr}d_{k,\sigma}^\yd d_{k,\sigma}^\nd t}.
\end{align}
Under this transformation, the operators are transformed as follows:
\begin{align}
\label{eq:op_trans}
U(t)c_1U^\yd(t)=&c_1,\\
\nonumber
U(t)c_2U^\yd(t)=&c_2e^{-i\frac{\omega_L-\omega_R}2t},\\
\nonumber
U(t)c_3U^\yd(t)=&c_3e^{-i\omega_{dr}t},\\
\nonumber
U(t)c_4U^\yd(t)=&c_4e^{-i\omega_Lt},\\
\nonumber
U(t)d_{k,\sigma}U^\yd(t)=&d_{k,\sigma}e^{-i\omega_{dr}t}.
\end{align}
The time derivative of $U(t)$ gives rise to an additional term in the Hamiltonian:
\begin{align}
\label{eq:u_der}
\begin{split}
\dot{U}(t)U^{-1}(t)=&
i\frac{\omega_L-\omega_R}2c^\yd_2 c^\nd_2
+i\omega^\nd_{dr}c^\yd_3c^\nd_3\\
&
+i\omega^\nd_Lc^\yd_4c^\nd_4
+\sum\limits_{k,\sigma}i\omega^\nd_{dr}d^\yd_{k,\sigma}d^\nd_{k,\sigma}
.
\end{split}
\end{align}
Plugging Eqs.~\eqref{eq:op_trans} and \eqref{eq:u_der} into Eq.~\eqref{eq:new_h}, we obtain:
\begin{align}
\begin{split}
H'_{\rm a}(t)=&\sum_{i=1}^4\Delta_i^\nd c^\yd_ic^\nd_i\\&+\Omega^\nd_L\left(c^\yd_1c^\nd_4+\mathrm{h.c.}\right)+\Omega^\nd_R\left(c^\yd_2c^\nd_3+\mathrm{h.c.}\right),
\end{split}\\
H'_{\rm b}(t)=&\sum\limits_{k,\sigma}\nu'^\nd_{k,\sigma}d^\yd_{k,\sigma}d^\nd_{k,\sigma},\\
\begin{split}
H'_{\rm ab,int}=&
\sum\limits_k\lambda^\nd_L\left(
d^\yd_{k,L}c^\yd_2c^\nd_4+d^\nd_{k,L}c^\yd_4c^\nd_2
\right)\\
&+\sum\limits_k\lambda_L\left(
e^{-2i\omega_{dr}t}d^\nd_{k,L}c^\yd_2c^\nd_4
+e^{2i\omega_{dr}t}d^\yd_{k,L}c^\yd_4c^\nd_2
\right)\\
&+\sum\limits_k\lambda^\nd_R\left(
d^\yd_{k,R}c^\yd_1c^\nd_3+d^\nd_{k,R}c^\yd_3c^\nd_1
\right)\\
&+\sum\limits_k\lambda_R\left(
e^{-2i\omega_{dr}t}d^\nd_{k,R}c^\yd_1c^\nd_3
+e^{2i\omega_{dr}t}d^\yd_{k,R}c^\yd_3c^\nd_1
\right).
\end{split}
\end{align}
This concludes the first step of the RWA.
We now move to the second step: neglecting the fast oscillating terms
that appear in the interaction part between the bath and the atom, $H'_{ab,int}$.
These terms oscillate at twice the driving frequency, $2\omega_{dr}$.
Because $\omega_{dr}$ is the largest energy scale in the problem, we can neglect these terms and obtain Eq.~\eqref{eq:rwa_haml}.
\section{Integrating degrees of freedom}
\label{sec:app5}
In this Appendix we give detail on the integration of rapidly oscillating degrees of freedom used in section \ref{sec:lowenegy}.
This operation, also known as ``adiabatic elimination'', can be preformed by converting  a given Hamiltonian to an action, and integrating the selected degrees of freedom.

The Hamiltonian we use to exemplify this procedure is given by:
\begin{align}
\label{eq:df_h}
H=&
\varepsilon^\nd_1c^\yd_1c^\nd_1+
g^\nd_{12}\left(c^\yd_1c^\nd_2+\mathrm{h.c.}\right)+g^\nd_{13}\left(c^\yd_1c^\nd_3+\mathrm{h.c.}\right),
\end{align}
where $c^{\left(\yd\right)}_i$ is the annihilation (creation) operator for the fermion $i$ and $g_{ij}$ is the coupling between the fermion $i$ and $j$.
This Hamiltonian has exactly the same structure as the one used in this manuscript (see Eq.~\eqref{eq:rwa_haml}).

We now show how to adiabatically eliminate the fermion $c_1$ and obtain an effective coupling between the fermion $c_2$ and $c_3$.
First, we use the Legendre transformation to compute the Lagrangian of Eq.~\eqref{eq:df_h}:
\begin{align}
\label{eq:df_l}
\begin{split}
L(t)=&\sum\limits_{i=1}^3ic^\yd_i\dot{c}^\nd_i-
\varepsilon^\nd_1c^\yd_1c^\nd_1\\&-
g^\nd_{12}\left(c^\yd_1c^\nd_2+\mathrm{h.c.}\right)-g^\nd_{13}\left(c^\yd_1c^\nd_3+\mathrm{h.c.}\right).
\end{split}
\end{align}
Next, we use Keldysh path integrals to integrate out the $c^\nd_1$ degree of freedom from the partition function:
\begin{align}
\label{eq:df_Z}
Z=&\int\mathcal{D}{\left[c^\nd_i,c^\yd_i\right]}e^{i\int_C\,L(t)}.
\end{align}
To achieve this goal, we first move to the Fourier domain and rewrite Eq.~\eqref{eq:df_l} as
\begin{align}
\label{eq:df_Lo}
\begin{split}
L(\omega)=&\sum\limits_{i=2}^3\omega c^\yd_ic^\nd_i+c^\yd_1\left(\omega-\varepsilon^\nd_1\right)c^\nd_1\\&-
\left(
c^\yd_1
\left(
g^\nd_{12}c^\nd_2+g^\nd_{13}c^\nd_3
\right)+\mathrm{h.c.}
\right).
\end{split}
\end{align}
Note that this expression has a form analogous to Eq.~\eqref{eq:Beresin}.
Let us define $G^{-1}_0=\omega-\varepsilon^\nd_1$ and a new operator
$V=g^\nd_{12}c^\nd_2+g^\nd_{13}c^\nd_3$, such that we can split the Lagrangian to three parts:
\begin{align}
L(\omega)=&L_1(\omega)-L_{\rm int}(\omega)+L_{23}(\omega),\\
L_1(\omega)=&c^\yd_1G^{-1}_0c^\nd_1,\\
L_{\rm int}(\omega)=&c^\yd_1V+\mathrm{h.c.},\\
L_{23}(\omega)=&\sum\limits_{i=2}^3\omega c^\yd_ic^\nd_i,
\end{align}
in order to integrate $c_1$ out, we ``complete the square'' of the sum of $L_1(\omega)$ and $L_{\rm int}(\omega)$:
\begin{multline}
L_1(\omega)+L_{\rm int}(\omega)=\\\left(c^\yd_1-V^\yd G^\nd_0\right)G^{-1}_0\left(c^\nd_1-G^\nd_0V^\nd\right)+V^\yd G^\nd_0V^\nd,
\end{multline}
converting the partition function to the following:
\begin{align}
\label{eq:df_Loe}
\begin{split}
Z=&
\int\mathcal{D}\left[c^\nd_i,c^\yd_i\right]\exp\left(i\int_C\,L_1(\omega)+L_{\rm int}(\omega)\right)
\\\times&
\exp\left(i\int_C\,L_{23}(\omega)\right)
\end{split}
\end{align}
Eq.~\eqref{eq:df_Loe} allows us to use the shifted Gaussian integral identity (see Eq.\eqref{eq:Beresin}) and obtain
\begin{align}
\begin{split}
Z\approx&\int\mathcal{D}\left[c^\nd_i,c^\yd_i\right]\exp\left(i\int_C\,V^\yd G_0V\right)\exp\left(i\int_C\,L_{23}(\omega)\right)\\
\approx&\int\mathcal{D}\left[c^\nd_i,c^\yd_i\right]\exp\left(i\int_C\,V^\yd G_0V+L_{23}(\omega)\right)\\
\approx&\int\mathcal{D}\left[c^\nd_i,c^\yd_i\right]\exp\left(i\int_C\,L_{\rm eff}(\omega)\right).
\end{split}
\end{align}
Here we define a new Lagrangian $L_{\rm eff}(\omega)=L_{23}(\omega)+V^\yd G_0V$.
We also note, that the definition of $G_0$ matches the one of Eq.~\eqref{eq:Gfbare}.
By making the same assumption, $\omega\ll\varepsilon_1$, the new effective Lagrangian is given by:
\begin{align}
L_{\rm eff}(\omega)=&\sum\limits_{i=2}^3\omega c^\yd_ic^\nd_i
-
\frac{g^\nd_{12}g^\nd_{13}}{\varepsilon^\nd_1}
\left(
c^\yd_2c^\nd_3+\mathrm{h.c.}
\right)
,
\\
L_{\rm eff}(t)=&\sum\limits_{i=2}^3ic^\yd_i\dot{c}^\nd_i-\frac{g^\nd_{12}g^\nd_{13}}{\varepsilon^\nd_1}\left(c^\yd_3c^\nd_2+\mathrm{h.c.}\right).
\end{align}
Using the Legendre transformation again we obtain:
\begin{align}
H_{\rm eff}(t)=&\frac{g^\nd_{12}g^\nd_{13}}{\varepsilon^\nd_1}\left(c^\yd_3c^\nd_2+\mathrm{h.c.}\right).
\end{align}
In summary, by integrating out the fermion 1, we obtained an effective model with direct coupling between fermion 2 and 3.

Using this method, one can transform Eqs.~\eqref{eq:rwa_haml} into Eqs.~\eqref{eq:Ham3}.

\section{Non-equilibrium dynamics of the generalized Dicke model}
\label{sec:app3}
\rm
In this Appendix, we provide the equations describing the real time dynamics of the generalized Dicke model (\ref{eq:gDM}-\ref{eq:nonumber}) with atomic and cavity dissipations.

\subsection{Green's functions}
\paragraph{$f$-fermion Green's function. --}
As introduced in the main text, the lesser, greater, and retarded Green's function of the $f$-fermion read
\begin{eqnarray}
G_{f}^{<}\left(t,t'\right)&=& i\av{ \hat f^\dag(t') \hat f(t)}, \\ \nonumber
G_{f}^{>}\left(t,t'\right) &=& - i\av{ \hat f(t) \hat f^\dag(t')}, \\ \nonumber
G_{f}^{R}\left(t,t'\right) &=& - i\theta(t-t')\av{ \lbrace \hat f(t), \hat f^\dag(t') \rbrace }.
\end{eqnarray}

Under the bare Hamiltonian $H_{0}$~\eqref{eq:gDM} the fermionic operators in the interaction picture evolve as $f\left(t\right)=f\left(0\right)e^{-i\omega_{0}t}$. This gives us the bare Green's functions
\begin{eqnarray}
G_{f,0}^{<}\left(t,t'\right) & = & i \cos\left( \omega_0 (t-t')\right) - i n_f(0) e^{i\omega_{0}\left(t-t'\right)}, \\ \nonumber
G_{f,0}^{>}\left(t,t'\right) & = & - \sin\left( \omega_0 (t-t')\right) -  i n_f(0) e^{i\omega_{0}\left(t-t'\right)}, \\ \nonumber
G_{f,0}^{R}\left(t,t'\right)  &=& - i\theta(t-t') e^{-i \omega_0 (t-t')}
\end{eqnarray}
At time $t'=t=0$, these equations set the initial conditions for the time evolution of the corresponding Green's functions.

In the dynamics, only two Green's function on the Keldysh contour are linearly independent, e.g., the lesser $G_f^<(t,t')$ and the retarded $ G_f^R(t,t')$ Green's functions. We can express the greater one using the following identity:
\begin{eqnarray}
G_f^>(t,t') &=& G_f^<(t,t') + G_f^R(t,t') - \left(G_f^R(t',t) \right)^\dag. \\ \nonumber
\end{eqnarray}

\paragraph{$\eta$-fermion Green's function. -- }
Majorana fermions satisfy the condition $\left\{ \eta\left(t\right),\eta\left(t'\right)\right\} =\left\{ \eta,\eta\right\} =2$. Thus, the Green's functions have the following form
\begin{eqnarray}
G_{\eta}^{<}\left(t,t'\right) & = & i\left\langle \eta\left(t'\right) \eta\left(t\right)\right\rangle, \\ \nonumber
G_{\eta}^{>}\left(t,t'\right) & = & -i\left\langle \eta\left(t\right)\eta\left(t'\right)\right\rangle, \\ \nonumber
G_{\eta}^{R}\left(t,t'\right) & = & -i\theta(t-t').
\end{eqnarray}
In the initial state, before coupling to the bath and cavity mode, the Green's function reads:
\begin{eqnarray}
&& G_{\eta,0}^<\left(t,t'\right)=i, \;\;\; G_{\eta,0}^>\left(t,t'\right)=-i \\ \nonumber
&& G_{\eta,0}^{R}\left(t,t'\right) = -i\theta(t-t').
\end{eqnarray}

One Majorana Green's function is independent and contains information about physical properties of the system, e.g., the greater one $G_\eta ^>(t,t') $. The lesser Green's function is
\begin{equation}
G_\eta ^<(t',t) =-G_\eta ^>(t,t').
\end{equation}

\paragraph{Cavity photon's Green's function. --}
When considering the solution of the generalized Dicke model, we should keep track of the anomalous terms of the cavity photon's Green's function.
As in the main text, we describe this Green's function with a $4\times4$ matrix in Keldysh-Nambu space. The lesser and retarded Green's functions are defined as
\begin{eqnarray}
D_{a}^{<}\left(t,t'\right)&=& -i\left(\begin{array}{cc}
\av{ \hat a^\dag(t')\hat a(t) }  & \av{ \hat a(t')\hat a(t) }  \\
\av{ \hat a^\dag(t') \hat a^\dag(t) }   &  \av{ \hat a(t')\hat a^\dag(t) }
\end{array}\right), \\ \nonumber
D_{a}^{R}\left(t,t'\right)&=&i \theta (t-t') \left(\begin{array}{cc}
\av{ \left[ \hat a(t),\hat a(t')^\dag\right] }  & \av{ \left[ \hat a(t),\hat a(t')\right] }  \\
\av{ \left[ \hat a^\dag(t),\hat a^\dag(t')\right]}   &\av{ \left[ \hat a^\dag(t),\hat a(t')\right] }
\end{array}\right).
\end{eqnarray}
The bosonic operators in the interaction picture evolve as $\left(a\left(t\right)=a\left(0\right)e^{i\omega_{c}t}\right)$.
This gives us the bare Green's functions
\begin{eqnarray}
D_{a,0}^{R}\left(t,t'\right)&=&i \theta (t-t') \left(\begin{array}{cc}
e^{ i \omega_c (t-t')} & 0   \\
0  & e^{ i \omega_c (t'-t)}
\end{array}\right), \\ \nonumber
D_{a,0}^{<}\left(t,t'\right)&&= - i \left(\begin{array}{cc}
n_a(0) e^{ i \omega_c (t'-t)} & 0\\
0 &  \left( n_a(0) + 1 \right) e^{ i \omega_c (t-t')}
\end{array}\right).
\end{eqnarray}

\paragraph{Green's functions of the dissipative bath. --}
Following the definitions from Appendix~\ref{sec:app1} for the greater and lesser Green's functions of the dissipative baths~\eqref{eq:bath_properties_app} in frequency domain, the corresponding Green's functions in real time read
\begin{eqnarray} \label{eq:bath_real_time}
&& \sum_{k}\frac{ i\Omega_L^2 \lambda_{k,L}^{2} }{ 8 \Delta_L^2 }  D_{k,L}^{>}(t,t')	 =    \gamma_\uparrow \delta(t-t'), \\ \nonumber
&& \sum_{k}\frac{i \Omega_L^2 \lambda_{k,L}^{2} }{ 8 \Delta_L^2 }  D_{k,L}^{<}(t,t') 	= 0, \\ \nonumber
&& \sum_{k}\frac{ i\Omega_R^2 \lambda_{k,R}^{2} }{ 8 \Delta_R^2 } D_{k,R}^{>}(t,t')    =    \gamma_\downarrow \delta(t-t'),\\ \nonumber
&& \sum_{k}\frac{ i\Omega_R^2 \lambda_{k,R}^{2} }{ 8 \Delta_R^2 } D_{k,R}^{<}(t,t')    = 0.
\end{eqnarray}

\subsection{Kadanoff-Byam equations}
Inverting the bare Green's function using the Leibnitz rule, we obtain the equations of motion for the retarded and lesser Green's functions
\begin{widetext}
\begin{eqnarray} \label{eq:noneq_dyn}
\left(i\frac{\partial}{\partial t} -\omega_{0} \right)G_{f}^{R}\left(t,t'\right) &=& \delta\left(t-t'\right)
+\int_{t'}^{t} ds \;\Sigma_{f}^{R}\left(t,s\right)G_{f}^{R}\left(s,t'\right) \\ \nonumber
\left( i\frac{\partial}{\partial t} -\omega_{0}\right) G_{f}^{<}\left(t,t'\right) &=& \int_{0}^{t}ds\; \Sigma_{f}^{R}\left(t,s\right)G_{f}^{<}\left(s,t'\right)
+\int_{0}^{t'}ds \;\Sigma_{f}^{<}\left(t,s\right)\left(G_{f}^{R}\left(t',s\right)\right)^{\dagger}\\\nonumber
\left(i\frac{\partial}{\partial t} \right) G_{\eta}^{R}\left(t,t'\right) &=& \delta\left(t-t'\right)
+\int_{t'}^{t}ds\; \Sigma_{\eta}^{R}\left(t,s\right)G_{\eta}^{R}\left(s,t'\right)  \\ \nonumber
\left( i\frac{\partial}{\partial t}\right) G_{\eta}^{<}\left(t,t'\right) &=& \int_{0}^{t}ds \;\Sigma_{\eta}^{R}\left(t,s\right)G_{\eta}^{<}\left(s,t'\right)
+\int_{0}^{t'}ds \;\Sigma_{\eta}^{<}\left(t,s\right)\left(G_{\eta}^{R}\left(t',s\right)\right)^{\dagger}\\ \nonumber
\left( \left( i\frac{\partial}{\partial t} + i \kappa \right) \sigma_z -\omega_c \right) D_a^R\left(t,t'\right) &=& \delta\left(t-t'\right)
+\int_{t'}^{t}ds \;\Pi_{a}^{R}\left(t,s\right)D_{a}^{R}\left(s,t'\right) \\ \nonumber
\left( \left( i\frac{\partial}{\partial t} + i \kappa \right) \sigma_z-\omega_c \right)  D_a^<\left(t,t'\right) &=&
\int_{0}^{t} ds\; \Pi_{a}^{R}\left(t,s\right)D_{a}^{<}\left(s,t'\right)
+\int_{0}^{t'}ds \;\Pi_{a}^{<}\left(t,s\right)\left(D_{a}^{R}\left(t',s\right)\right)^{\dagger}
\end{eqnarray}
\end{widetext}
where $\sigma_z$ is the Pauli matrix.

We define the self-energies using the diagrams shown in Fig.~\ref{fig:Dicke_diagrams} in the main text.
In the leading order, the self-energy of the $f$- and $\eta$-fermions contain only the contribution proportional to the Green's functions of the dissipative bath.
For the $f$-fermion we obtain the following expressions
\begin{eqnarray}\label{eq:sigma_f}
\Sigma_f^{<}(t,t') &=& \frac{i}{2} \sum_k \frac{\Omega_L^2}{\Delta_L^2} \lambda_k^2  D_{k,L}^{<} (t,t') G_\eta^<(t,t') \\ \nonumber
&+&\frac{i}{2} \sum_k \frac{\Omega_R^2}{\Delta_R^2} \lambda_k^2  D_{k,R}^{>} (t,t') G_\eta^<(t,t'), \\ \nonumber
\Sigma_f^{>}(t,t') &=& \frac{i}{2} \sum_k \frac{\Omega_L^2}{\Delta_L^2} \lambda_k^2  D_{k,L}^{>} (t,t')  G_\eta^>(t,t')  \\ \nonumber
&+&  \frac{i}{2} \sum_k \frac{\Omega_R^2}{\Delta_R^2} \lambda_k^2  D_{k,R}^{<} (t,t')  G_\eta^>(t,t')
\end{eqnarray}
Substituting Eq.~\eqref{eq:bath_real_time}, equations~\eqref{eq:sigma_f} simplify to
\begin{eqnarray}
\Sigma_f^{<}(t,t') &=&  \gammau \delta(t-t') G_\eta^<(t,t') \\ \nonumber
 \Sigma_f^{R}(t,t') &=&  -\left( \gammad + \gammau\right)\delta(t-t')   G_\eta^<(t,t').
\end{eqnarray}
Note that we didn't use Nambu notation to describe the $f$-fermion. In the leading order, the Green's function of the $f$-fermion doesn't acquire any anomalous terms in the transient dynamics.

Similarly, we calculate the self-energies of the $\eta$-fermion
\begin{eqnarray}\label{eq:sigma_eta}
\Sigma_\eta^{<}(t,t') &=&  i \sum_k \frac{\Omega_R^2}{\Delta_R^2} \lambda_k^2  D_{k,R}^{<} (t,t')  G_f^<(t,t')  \\ \nonumber
&+& i \sum_k \frac{\Omega_L^2}{\Delta_L^2} \lambda_k^2  D_{k,L}^{>} (t,t')  G_f^<(t,t'),  \\ \nonumber
\Sigma_\eta^{>}(t,t') &=& i \sum_k \frac{\Omega_R^2}{\Delta_R^2} \lambda_k^2  D_{k,R}^{>} (t,t') G_f^>(t,t')\\ \nonumber
&+&  i \sum_k \frac{\Omega_L^2}{\Delta_L^2} \lambda_k^2  D_{k,L}^{<} (t,t') G_f^>(t,t').
\end{eqnarray}
Substituting Eq.~\eqref{eq:bath_real_time}, we obtain
\begin{eqnarray}\label{eq:selfen_1}
\Sigma_\eta^{<}(t,t') &=& 2\gammad \delta(t-t')  G_f^<(t,t')  \\ \nonumber
 \Sigma_\eta^{R}(t,t') &=& 2 \gammau \delta(t-t')  G_f^>(t,t') - 2 \gammad \delta(t-t')  G_f^<(t,t').
\end{eqnarray}

By analogy with the cavity photon's polarization operator in the steady state~\eqref{eq:pi_gfgeta}, we have the following expressions of the polarization operator in the transient dynamics:
\begin{widetext}
\begin{eqnarray}
	\label{eq:selfen_2}
&&\Pi_a^{<}(t,t') = \frac{i}{2} \Lambda^T \arr4{  G_f^<(t,t') G_\eta^>(t',t)}{0}{0}{ G_\eta^<(t,t') G_f^>(t',t)} \Lambda\\ \nonumber
&&\Pi_a^{R}(t,t') = \frac{i}{2} \Lambda^T \arr4{  G_f^<(t,t')  G_\eta^A(t',t) + G_f^R(t,t')  G_\eta^<(t',t)}{0}{0}{G_\eta^<(t,t')  G_f^A(t',t) + G_\eta^R(t,t')  G_f^<(t',t)} \Lambda
\end{eqnarray}
\end{widetext}
where $\Lambda$ is the interaction vertex given by the Eq.~\eqref{eq:Nambu}.
Solution of equations~\eqref{eq:noneq_dyn} with self-energies~\eqref{eq:selfen_1} and \eqref{eq:selfen_2}  will describe the dynamics of the generalized Dicke model after instantaneous coupling to the bath and cavity modes.

\end{appendix}

\end{document}